\documentclass[12pt]{iopart}
\usepackage{iopams}
\usepackage{graphicx}
\usepackage{psfig}
\begin{document}

\title[Regional Averaging and Scaling in Relativistic Cosmology]
{Regional Averaging and Scaling \\ in Relativistic Cosmology}

\medskip

\author{Thomas Buchert$^a$ and Mauro Carfora$^b$}

\smallskip

\address{$^a$
Theoretische Physik, Ludwig--Maximilians--Universit\"at,
Theresienstr. 37, 80333 M\"unchen, Germany. 
Email: buchert@theorie.physik.uni-muenchen.de}

\smallskip

\address{$^b$
Dipartimento di Fisica Nucleare e Teorica,
Universit\`{a} degli Studi di Pavia
and Istituto Nazionale di Fisica Nucleare, Sezione di Pavia,
via A. Bassi 6, 27100 Pavia, Italy.
Email: mauro.carfora@pv.infn.it}

\begin{abstract}
Averaged inhomogeneous cosmologies lie at the forefront of interest, since 
cosmological parameters such as the rate of expansion or the mass density are
to be considered as volume--averaged quantities and only these can be compared
with observations. For this reason the relevant parameters are intrinsically 
scale--dependent and one wishes to control this dependence without restricting
the cosmological model by unphysical assumptions. In the latter respect we 
contrast our way to approach the averaging problem in relativistic cosmology 
with shortcomings of averaged Newtonian models. Explicitly, we investigate the 
scale--dependence of Eulerian volume averages of scalar functions on Riemannian 
three--manifolds. We propose a complementary view of a Lagrangian smoothing of 
(tensorial) variables as opposed to their Eulerian averaging on spatial domains. 
This programme is realized with the help of a global Ricci deformation flow for 
the metric. We explain rigorously the origin of the Ricci flow which, on 
heuristic grounds, has already been suggested as a possible candidate for 
smoothing the initial data set for cosmological spacetimes. The smoothing of 
geometry implies a renormalization of averaged spatial variables. We discuss 
the results in terms of effective cosmological parameters that would be 
assigned to the smoothed cosmological spacetime. In particular, we find that 
the on the smoothed spatial domain $\overline{\cal B}$ evaluated cosmological
parameters obey $\overline{\Omega}_{\overline{\cal B}}^{m} + 
\overline{\Omega}_{\overline{\cal B}}^{R} + 
\overline{\Omega}_{\overline{\cal B}}^{\Lambda} + 
\overline{\Omega}_{\overline{\cal B}}^{\cal Q} = 1$,
where $\overline{\Omega}_{\overline{\cal B}}^m$, 
$\overline{\Omega}_{\overline{\cal B}}^R$ and 
$\overline{\Omega}_{\overline{\cal B}}^{\Lambda}$ correspond to the standard
Friedmannian parameters, while $\overline{\Omega}_{\overline{\cal B}}^{\cal Q}$ 
is a remnant of cosmic variance of expansion and shear fluctuations on the
averaging domain. All these parameters are `dressed' after smoothing--out 
the geometrical fluctuations, and we give the relations of the `dressed' to the 
`bare' parameters. While the former provide the framework of interpreting 
observations with a ``Friedmannian bias'', the latter determine the actual 
cosmological model.

\end{abstract}


\maketitle


\section*{Introduction}

Research on cosmological spacetimes has been in the realm of general 
relativity 
for a long time, establishing the standard cosmological models that are based 
on homogeneous 
(and mostly isotropic) solutions of Einstein's laws of gravity for a 
continuous fluid. 
Spatially homogeneous spacetimes are understood to a high degree, and 
cosmologies based upon them
certainly lie in a well--charted terrain. The difficulties or better 
challenges arise, if we want to respect
the actually present inhomogeneities in the Universe. Newtonian continuum 
mechanics 
appears to be a simpler theory to model the inhomogeneous Universe, usually 
restricted to 
the matter dominated epoch on subhorizon scales. Indeed, 
most contemporary efforts for the modelling of inhomogeneities are based on 
Newtonian cosmological models. 
There is, however, a drawback: Newtonian continuum mechanics of a 
self--gravitating 
fluid is not a proper theory {\em per se} \cite{ehlers:festschrift}, but has 
to be setup with
suitable boundary conditions; for the cosmological modelling it is practically 
restricted to setting up the physical variables relative to a
homogeneous background, while the (inhomogeneous) deviations thereof have to 
be subjected to periodic
boundary conditions. Even though we may accept periodic boundary conditions as 
a necessary cornerstone of a Newtonian model -- hence, we view the Universe as 
a caleidoscope of ever--repeating
self--similar boxes that are supposed to supply a `fair sample' -- the 
introduction of a global reference 
background is essential to do so. This may be illustrated as follows. Consider 
Poisson's equation for the Newtonian gravitational potential. This potential cannot
be periodic as a whole (e.g. for a homogeneous background it grows 
quadratically with 
distance from an origin). Moreover, solutions of Poisson's equation are only unique, 
if the spatial average of the source vanishes. Both requirements, periodicity 
and uniqueness, can be accomplished only for fields defined as inhomogeneous 
deviations from a given reference background, e.g. the standard FLRW models 
(for details compare \cite{buchert:average}). 
Note that most currently employed models including numerical N--body 
simulations rest on these assumptions.
These ``forcing conditions'' must be considered a drawback for the following 
reason: 
we may consider the spatially averaged variables as replacing the former 
homogeneous variables, e.g., the 
volume--averaged rate of expansion measuring the Hubble law on a given 
averaging scale. This (effective) expansion rate
does not obey the Friedmann equations of the
standard cosmological models, but the true equation features an additional 
source term 
due to kinematical fluctuations \cite{buchert:average}.  
As was first pointed out by Ellis \cite{ellis:relativistic}, this so--called 
``backreaction effect'' is a result of the 
nonlinearity of the basic system of equations, if general relativity or 
Newtonian theory, lying at the heart of the problem how we could compare and match
the FLRW standard model of cosmology with an averaged inhomogeneous model 
\cite{ellisstoeger}.
It is here, were the restriction of using a Newtonian model becomes evident: 
this extra source term is a full divergence of a vector 
field and, hence, consistently vanishes on the periodic boundary. It is, 
however, {\em not} 
a full (three--dimensional) divergence on non--Euclidean spaces (see 
\cite{buchert:grgdust}
for a discussion of this point). 
Hence, although it is commonly agreed that observables like Hubble's 
``constant'' or the
mass density depend on the surveyed volume of space and must be intrinsically 
scale--dependent, the Newtonian models
have to introduce a ``largest scale'' where these observables assume a 
constant value
that is determined by the homogeneous standard cosmology and, consequently, by 
initial conditions given on the 
largest scale only. {\em By construction}\footnote{Cosmologists were employing
this construction for a long time without justification, and they have been
``lucky'' that the backreaction term is indeed a full divergence which in turn
implies that it vanishes on the periodic simulation box comoving with a 
standard Hubble flow. Without this (non--trivial) property of the backreaction
term cosmological N--body simulations would just be artificial constructions.}
an averaged Newtonian model features the 
characteristics of the standard homogeneous models. According to what has been 
said above,
this scale appears artificial and we may not have such a scale where the 
averaged inhomogeneous model
can be identified with or approximated {\em for all times} by the standard 
model.
In other words, we may not find a global frame comoving with a standard 
Hubble flow and at the same time providing the evolution on average.
As an example we point out that the Newtonian curvature parameter is 
determined by
the initial data on the periodicity scale (e.g., a flat Einstein--de Sitter 
cosmology remains so during the 
evolution), while in general relativity the averaged scalar curvature is 
coupled to the
backreaction of the inhomogeneities \cite{buchert:grgdust}. Since the 
dynamical evolution of the curvature 
parameter on scales smaller than the periodicity scale strongly depends on the 
inhomogeneities (see \cite{buchert:bks} for a quantitative investigation), we 
can expect that a generic averaged cosmology will not 
keep the global average curvature at this initial value; it will, like any 
other
variable, change in the coarse of evolution (for further discussion
see \cite{buchert:onaverage}). 
It should be remarked here that very often the argument is advanced that the
backreaction term is negligible, because it is numerically small. On sufficiently large scales
the latter is supported by some of the following investigations:
\cite{futamase1}, \cite{bildhauerfutamase1},
\cite{bildhauerfutamase2}, \cite{seljakhui}, \cite{futamase2}, \cite{mukhanov:backreaction},
\cite{russetal:backreaction}, \cite{buchert:bks}. Still, a small perturbation
can (and as shown in \cite{buchert:bks}) will drive the dynamical system for
the averaged fields into another ``basin of attraction'' implying drastic
changes of the volume--averaged cosmological parameters {\it although} the
backreaction term is numerically small.

This ``kinematical backreaction'' representing, roughly speaking, the influence of 
fluctuations in the matter fields on the effective (spatially averaged)
dynamical properties of a spatial region in the Universe, does not comprise
the whole story. Even if we take the influence of fluctuations on the
averaged variables into account, these variables themselves still depend on 
the bumpy geometry of the 
inhomogeneous averaging region. It turns out that this problem is quite
subtle and lies
at the heart of any interpretation of observables in terms of a cosmological
model: observed average characteristics of a surveyed region are, by lack of
better standards, taken as averages on a Euclidean (or constant curvature) 
space section. Any matter averaging program in relativistic cosmology is not 
complete unless we also devise a way to interprete the averages on an
averaged geometry. The latter, however, is a tensorial entity for which 
unique procedures of averaging are not at hand. 
In the present paper we especially address this problem and propose a 
Lagrangian smoothing of tensorial variables
as opposed to their Eulerian averaging.   
The present investigation will reveal a further shortcoming of Newtonian 
cosmology: curvature fluctuations turn out to be crucial and may even outperform
the effect of kinematical fluctuations quantitatively.  

In summary we can state the following headline of our investigation
of the averaging problem: since averaged scalar characteristics form an 
important set of parameters
that respectively constrain or are determined by observations, it is a highly 
relevant task to develop a theoretical framework for
averaging and scaling within which the currently collected datasets can be 
analyzed reliably and free of 
unphysical model assumptions. Newtonian models are, due to their very 
architecture, 
not free of such assumptions.
In light of this the modelling of cosmologies 
has to be lifted back on the stage of general relativity, leaving behind the
Newtonian ``toy universe models'', which 
were helpful to understand basic properties of the formation of structure, but 
have reached a dead end where the Euclidean periodic box is taken for real and every 
observational data  ``fitted'' to its parameters.
Fortunately, the structure of the basic equations that govern the averages of
observables  
in general relativity is so close to
their Newtonian counterparts, that it is evident to better work in the 
relativistic framework (compare the formal equivalence of the 
effective expansion law in Newtonian cosmology \cite{buchert:average} with 
that in general relativity \cite{buchert:grgdust}).
The challenge that we shall meet here, and the answer that we shall provide,
concerns the interpretation of the average characteristics in a regional
survey within a smoothed--out cosmological spacetime.

\medskip
      
Let us now come to the content of this article and describe our approach
before we formalize it.
Consider a three--dimensional manifold equipped with a Riemannian metric 
$(\Sigma,g_{ab})$.
On such a hypersurface we may select a simply--connected spatial region
and evaluate certain average properties of the 
physical (scalar) variables on that domain such as, e.g., the
volume--averaged density field, or the volume--averaged scalar curvature. 
These (covariant) average values are functionally dependent on position 
and the geometry of the chosen domain of averaging.
Let us now be more specific and relate the averages to scaling properties of 
the 
physical variables. It is then natural to identify the domain with a geodesic 
ball
centred on a given position, and introduce the spatial scale via the geodesic 
radius
of the ball. We may consider the variation of this radius and so 
explore the scaling properties of average characteristics on the whole 
manifold. 
It turns out that this scaling depends on the intrinsic geometry of the 
hypersurface, and we shall explicitly evaluate this dependence in the neighborhood
of the domain of averaging. 
Averaging regionally at every position chosen in the spatial slice at a fixed
averaging scale, we arrive at averaged fields 
that, upon changing the scale, depend on the accidents of the regional
geometry. 
   
We may call this point of view Eulerian -- and everybody would first think of 
this point of view -- 
in the sense that the spatial manifold is explored {\em passively} by blowing 
up the geodesic 
balls covering larger and larger volumes with a larger amount of material
mass. On the contrary, the key idea of our 
approach consists in demonstrating 
that a corresponding {\em active} averaging procedure can be devised which is 
Lagrangian: 
we hold the ball at a fixed (Lagrangian) radius, and  
deform the dynamical variables (actively) inside the balls such that the 
deformation corresponds to the
smoothing of the fields. The (first) variation with scale of, e.g.,
the density field or the metric is then mirrored by one--parameter families of 
successively 
deformed density fields and metrics. We shall show that this deformation 
corresponds to a first variation of the metric along the Ricci tensor, known 
as the Ricci
flow. Since this flow has received great attention in the mathematical 
literature
-- major contributions are due to Hamilton \cite{hamilton:ricciflow1}, 
\cite{hamilton:ricciflow2} -- we shall so translate the averaging 
procedure into a well--studied deformation flow, and we shall do it much in 
the spirit of a 
renormalization group flow \cite{carfora:deformation1}, 
\cite{carfora:deformation2}, \cite{carfora:RG}, \cite{hosoya:RG}.
As is common practice in the literature on the Ricci flow, 
we adopt the normalization that this flow is globally 
volume--preserving\footnote{We have shown that a homothetic transformation of the variables 
would allow for a different normalization, e.g. such that the Ricci flow 
is globally mass--preserving. This, however, plays no significant role in our 
investigation of the regional averaging; the result is strictly equivalent.}.  

We prefer the Lagrangian point of view also for reasons of the following wider 
perspective.
Suppose that the chosen hypersurface is a member of a given foliation of 
spacetime and,
as an illustration, let the dynamical flow be geodesic.
Then, the Einstein dynamics of, e.g., the spatial metric $g_{ab}$ is a first 
variation (in 
proper time $t$) of the metric in the direction of the extrinsic curvature 
tensor 
$K_{ab}$, $\partial_t g_{ab} = -2 K_{ab}$. Evaluating the dynamical
properties on spatial domains (geodesic balls)  with fixed Lagrangian radius 
amounts to
following the motion of the collection of fluid elements inside the ball along 
their spacetime trajectories, 
thus keeping the number of fluid 
elements (and hence their  total mass) fixed during the evolution. This 
program has been carried out for
the matter models `irrotational dust', `irrotational perfect fluid' and 
`scalar field' 
in \cite{buchert:grgdust}, \cite{buchert:grgfluid},
\cite{buchert:grgscalarfield}. Complementary to this time--evolution picture (in the
direction of the extrinsic  curvature), we shall here investigate a
d`Alembertian   ``virtual evolution'' (first variation in the direction of the
Ricci  tensor ${\cal R}_{ab}$ with variation parameter $\beta$), where also in
this case the material mass in the domain of averaging is kept constant. 
Our headline of forthcoming work is to setup a
renormalized effective dynamics that includes spatial and temporal variation
simultaneously.  


\newpage

\section{Averaging and Scaling put into Perspective}

Addressing the issue of scaling in the Newtonian framework, one thinks of
scaling properties of the dependent and independent variables under 
a transformation of the spatial Eulerian coordinates of the form
$x^i \mapsto L x^i$ (and a suitable scaling for the time--variable), where
$L$ is the parameter of the scale transformation (see, e.g., \cite{sota:RG}).
Such a transformation may be used as a dimensional analysis and for the
purpose of constructing self--similar solutions of the Euler--Poisson system of
equations. In general relativity we cannot have such a naive rescaling
structure. Spacetime is a dynamic entity; as we shall see, rescaling of the
form $x^{i}\mapsto L x^{i}$ must be replaced by a point--dependent functional
rescaling, otherwise the scaled fields are strictly equivalent to the
original fields.

In order to characterize the correct conceptual framework for addressing
averaging and scaling properties in relativistic cosmology, let us first remark
that in general relativity we basically have one scaling variable related to
the unit of distance: we can express the unit of time in terms of the unit
of distance using the speed of light. Similarly, we can express the unit of
mass through the unit of distance using Newton's gravitational constant.
This remark implies that the scaling properties of Einstein's equations
typically generate a mapping between distinct initial data sets, (in this
sense we are basically dealing with a spatial renormalization group 
transformation).
As we shall see below in detail, a rigorous characterization of an averaging
procedure in relativistic cosmology is indeed strictly connected to the
scaling geometry of the initial data set for Einstein's equations. 

\begin{figure}
\begin{center}
\includegraphics[width=7cm]{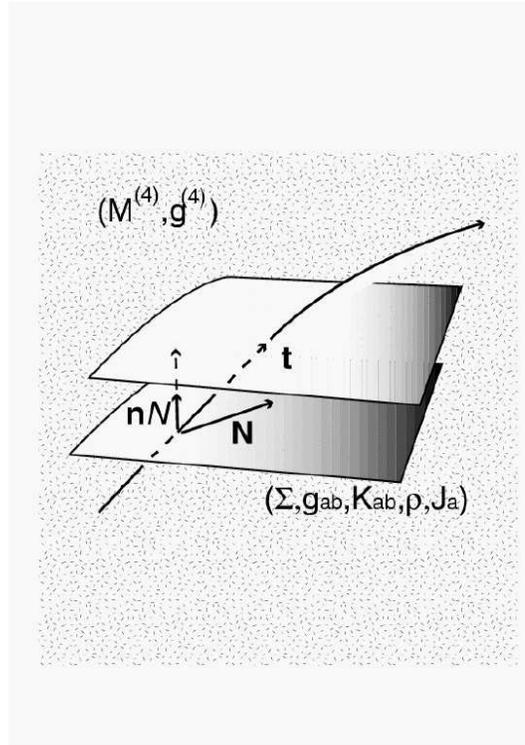}
\end{center}
\caption{\label{fig:adm} 
An admissible set of initial data is propagated along the time--like vector field
$\mathbf{t} = \mathbf{n} N + \mathbf{N}$, with the lapse function $N$ scaling the 
unit normal $\mathbf{n}$ at the given point and the shift vector field $\mathbf{N}$.
The spacetime geometry is foliated into hypersurfaces
$\Sigma$, $(M^{(4)}\simeq \Sigma \times \mathbb{R}, g^{(4)})$, on which we study scaling 
properties of the initial data set.
}
\end{figure}

Thus, it seems appropriate to recall some of the properties of the
Arnowitt--Deser--Misner formulation of Einstein's equations. This essentially
exploits the fact that a globally hyperbolic spacetime $(M^{(4)} ,
g^{(4)})$ is diffeomorphic to
$\mathbb{R}\times \Sigma $, where $\Sigma $ is a three--dimensional manifold
which we assume (for simplicity) to be closed. The explicit diffeomorphism
$\varphi :\mathbb{R}\times \Sigma \rightarrow  M^{(4)}$ is constructed by
defining on $M^{(4)}$ a timelike vector field $\mathbf{t}$ the integral curves
of which, $\Phi (t,p)$,  define a map $f:\Sigma \rightarrow M^{(4)}$ according
to $p\longmapsto f(p)=\Phi (t,p)$, (for each given $t\in \mathbb{R}$).
Consider, in such a setting, the dynamics of a self--gravitating distribution
of matter. In the ADM formulation, the dynamics of such a distribution and the
corresponding spacetime geometry $(M^{(4)}\simeq \Sigma \times \mathbb{R},
g^{(4)})$ are described by the evolution of an initial data set:
\begin{equation}
(\;\;\Sigma\;,\; g_{ab}\;,\;K_{ab}\;,\;\varrho\;,\;J_{a}\;\;)
\end{equation}
where the Riemannian metric $g_{ab}$ and the triple of tensor fields 
$(K_{ab},\varrho ,J_{a})$\footnote{Latin indices run through $1,2,3$; we adopt
the summation convention. The nabla operator denotes covariant derivative with 
respect to the 3--metric. The units are such that $c=1$.}, are
subjected to the Hamiltonian and divergence constraints:
\begin{eqnarray}
{\cal R}+{K}^{2}-K^a_{\;\,b}K^{b}_{\;\,a} &=&16\pi G\varrho +2\Lambda \;, \\
\nabla _{b}K^{b}_{\;\,a}-\nabla_{a}K &=& 8\pi G J_{a}\;,
\label{constraints}
\end{eqnarray}
where $K:= g^{ab}K_{ab}$ is the trace of the extrinsic curvature $K_{ab}$, 
${\cal R}:=g^{ab}{\cal R}_{ab}$ is the trace of the intrinsic Ricci curvature of the metric
$g_{ab}$, and $\Lambda $ is the cosmological constant. If such a set of
admissible data is propagated according to the evolutive part of 
Einstein's equations, then the symmetric tensor field $K_{ab}$ can be
interpreted as the second fundamental form of the embedding of 
$(\Sigma,g_{ab})$ 
in the spacetime $(M^{(4)}\simeq \Sigma \times \mathbb{R}, g^{(4)})$
resulting from the evolution of $(\Sigma ,g_{ab},K_{ab},\varrho ,J_{a})$,
whereas $\varrho $ and $J_{a}$ are, respectively, identified with the mass
density and the momentum density of the material self--gravitating sources on 
$(\Sigma ,g_{ab})$. Explicitly, the evolution equations associated with the
initial data $(\Sigma ,g_{ab},K_{ab},\varrho ,J_{a})$ are provided (in the
absence of stresses, \emph{i.e.}, for a dust matter model\footnote{
We restrict our consideration to the simplest matter model, although almost 
all following investigations would apply to more general matter models.}) by 
\begin{eqnarray}
\fl \frac{\partial g_{ab}}{\partial t}=-2NK_{ab}+{\cal L}_{\vec N}g_{ab}\;,\\
\fl \frac{\partial K_{ab}}{\partial t}=-\nabla _{a}\nabla _{b}N+N\left[ K
K_{ab}+{\cal R}_{ab}-(4\pi G\varrho +\Lambda )g_{ab}\right] +{\cal L}_{\vec 
N}K_{ab}\;,
\end{eqnarray}
where $N$ and $\vec N$, respectively, denote the lapse function
and the shift vector field associated with the mapping $f:\Sigma \rightarrow 
M^{(4)}$ (\emph{i.e.}, $\vec t =N\vec n + \vec N$, 
$\vec n$ being the future--pointing unit
normal to the embedding $\Sigma \hookrightarrow M^{(4)}$ (Fig.~\ref{fig:adm})). 

\medskip

For the discussion of scaling properties and the averaging procedure
associated with an admissible set of initial data $(\Sigma,
g_{ab},K_{ab},\varrho ,J_{a})$ for a cosmological spacetime $(M^{(4)}\simeq
\Sigma \times \mathbb{R},g^{(4)})$ we start by characterizing
explicitly a scale--dependent averaging for the empirical mass distribution
$\varrho $.

\subsection{Interlude: matter seen at different scales}

\begin{figure}
\begin{center}
\includegraphics[width=7cm]{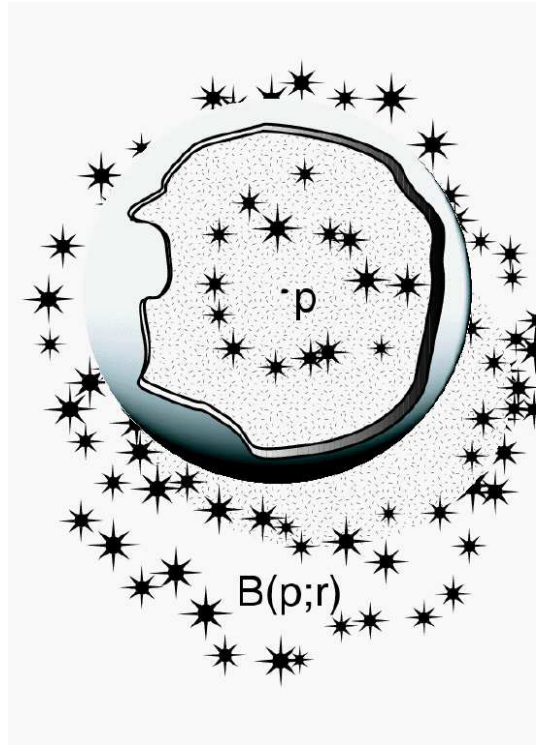}
\end{center}
\caption{\label{fig:averages} 
The domain of averaging, centred on a given point $p$, is a (possibly non--convex) simply--connected
geodesic ball domain contained in the hypersurface. The diagnostic parameter to explore the 
manifold is the radius $r$ of the geodesic ball. It is mapped from the ``master ball'' with 
Euclidean geometry into the manifold by the exponential mapping.  
The shape of the ball's surface reflects the inhomogeneous geometry of the hypersurface
at the scale $r$; the ball encloses a portion of the inhomogeneous matter distribution
(marked by stars) interacting with the environmental distribution. 
}
\end{figure}

In order to characterize the scale over which we are smoothing the empirical
mass distribution $\varrho $, we need to study the distribution $\varrho $ by
looking at its average behavior on regional domains (geodesic balls) 
in $(\Sigma ,g_{ab})$ with
different centers and radii. The idea is to move from the function $\varrho
:\Sigma \rightarrow \mathbb{R}^{+}$ to an associated function, defined on $%
\Sigma \times \mathbb{R}^{+}$, and which captures some aspect of the
behavior of the given $\varrho $ on average, at different scales and locations.
The simplest function of this type is provided by the {\em regional 
volume average}: 
\begin{equation}
\left\langle \varrho \right\rangle _{B(p;r)}:= 
\frac{1}{V\left(
B(p;r)\right) }\int_{B(p;r)}\varrho \;d\mu _{g},  \label{locav}
\end{equation}
where $p\in \Sigma $ is a generic point, $d\mu _{g}$ is the Riemannian
volume element associated with $(\Sigma ,g_{ab})$,\ and $B(p;r)$ denotes the
geodesic ball at center $p$ and radius $r$ in $(\Sigma ,g_{ab})$
(Fig.~\ref{fig:averages}), \emph{i.e.}, 
\begin{equation}
B(p;r):= \left\{ q\in (\Sigma ,g_{ab})\,:d_{g}(p,q)\leq r\,\,\right\} ,
\end{equation}
where $d_{g}(p,q)$ denotes the distance, in $(\Sigma ,g_{ab})$, between the 
point 
$p$ and $q$. Note that if 
\begin{equation}
diam:= \sup \left\{ d_{g}(p,q)\,:p,q\in (\Sigma ,g_{ab})\right\}
\end{equation}
denotes the diameter of $(\Sigma ,g_{ab})$, then as $r\rightarrow diam$, we 
get 
\begin{equation}
\left\langle \varrho \right\rangle _{B(p;r)}\longrightarrow \left\langle 
\varrho
\right\rangle _{\Sigma }:= \frac{1}{V(\Sigma)}%
\int_{\Sigma }\varrho \;d\mu _{g},
\end{equation}
at any point \bigskip\ $p\in \Sigma $. Conversely, if $\varrho :\Sigma
\rightarrow \mathbb{R}^{+}$ is locally summable, then 
\begin{equation}
\lim_{r\rightarrow 0}\,\left\langle \varrho \right\rangle _{B(p;r)}=\varrho (p)\;,
\end{equation}
for almost all points $p\in \Sigma $. The passage from $\varrho $ to $
\left\langle \varrho \right\rangle _{B(p;r)}$ corresponds to replacing the
position--dependent empirical distribution of matter in $B(p;r)$ by a
regionally uniform distribution $\left\langle \varrho \right\rangle _{B(p;r)}$
which is constant over the typical scale $r$. Note that for the total
(material) mass contained in $B(p;r)$, we get 
\begin{equation}
M(B(p;r)):= \int_{B(p;r)}\varrho \;d\mu _{g} = V\left( B(p;r)\right)
\left\langle \varrho \right\rangle _{B(p;r)},  \label{locmass}
\end{equation}
and 
\begin{equation}
\lim_{r\rightarrow diam}M(B(p;r))=M(\Sigma ):= \int_{\Sigma }\varrho \;d\mu _{g},
\end{equation}
where $M(\Sigma )$ is the total (material) mass contained in $(\Sigma 
,g_{ab})$.
Our expectations in $\left\langle \varrho \right\rangle _{B(p;r)}$ are
motivated by the fact that on Euclidean $3$--space $(\mathbb{R}
^{3},\delta _{ab})$, if $\varrho $ is a bounded function, then the regional
average $\left\langle \varrho \right\rangle _{B(p;r)}$ is a Lipschitz function
on $\mathbb{R}^{3}\times \mathbb{R}^{+}$, endowed with the hyperbolic 
metric \cite{semmes}:
\begin{equation}
\frac{dx_{1}^{2}+dx_{2}^{2}+dx_{3}^{2}+dr^{2}}{r^{2}}.
\end{equation}
In other words,

\begin{equation}
\left| \left\langle \varrho \right\rangle _{B(p;r)}-\left\langle \varrho
\right\rangle _{B(q;r)}\right| \leq C_{0}d_{h}(p,q),
\end{equation}
where $C_{0}$ is a constant and $d_{h}(p,q)$ is the (hyperbolic) distance
between $p$ and $q$. Thus, the regional averages $\left\langle \varrho
\right\rangle _{B(p;r)}$ do not oscillate too wildly 
as $p$ and $r$ vary,
and the replacement of $\varrho $ by $\{\left\langle \varrho \right\rangle
_{B(p;r)}\}_{p\in \mathbb{R}^{3}}$ indeed provides an averaging of the
original matter distribution over the length scale $r$. It is not obvious
that such a nice behavior carries over to the Riemannian manifold $(\Sigma
,g_{ab})$. The point is that, even if the regional averages $\{\left\langle 
\varrho
\right\rangle _{B(p;r)}\}_{p\in \Sigma }$ provide a controllable device 
of smoothing the matter distribution at the given scale $r$ (see: Remark 1), 
they still depend in a sensible way on the geometry of the typical ball
$B(p;r)$ as we vary the averaging radius. In this connection we need to
understand how, as we rescale the domain $B(p;r)$, the regional average
$\left\langle \varrho \right\rangle _{B(p;r)}$ depends on the underlying 
geometry of $(\Sigma ,g_{ab})$.
The reasoning here is slightly delicate, so we go into a few details that
require some geometric preliminaries.

\bigskip

{\sl Geometrical evolution of geodesic ball domains}

\begin{figure}
\begin{center}
\includegraphics[width=7cm]{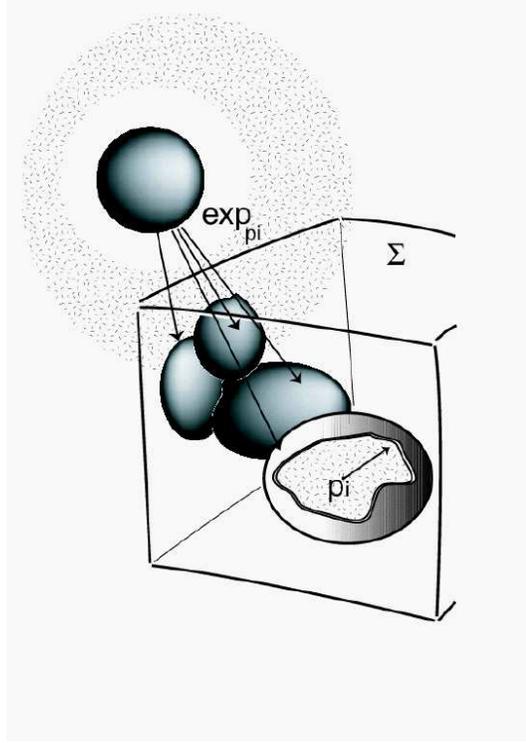}
\end{center}
\caption{\label{fig:balldomains} 
Four ball domains mapped by the exponential mapping from the Lagrangian 
``master ball'' into the Riemannian manifold are shown. Notice that the ball
domains may overlap. The possible case in which they lie in 
topologically disconnected pieces of the manifold is excluded due to our
choice of a maximal radius.}  
\end{figure}

Let us denote by 
\begin{eqnarray}
\exp _{p}:T_{p}\Sigma \rightarrow \Sigma \nonumber \\
(\vec{v},r)\longmapsto \exp _{p}(r\vec{v})
\label{expmap}
\end{eqnarray}
the exponential mapping at $p\in (\Sigma ,g_{ab})$, \emph{i.e.}, the map
which to the vector $r\vec{v}\in T_{p}\Sigma \simeq \mathbb{R}^{3}$
associates the point $\exp _{p}(r\vec{v})\in \Sigma $ reached at ``time'' 
$r\in \mathbb{R}_{+}$ by the unique geodesic issued at $p\in \Sigma $ with
unit speed $\vec{v}\in \mathbb{S}^{2}(1)$ . Let $r>0$ be such that $\exp
_{p} $ is defined on the Euclidean ball 
\begin{equation}
B_{E}(0,r):= \{y\in T_{p}\Sigma \simeq \mathbb{R}^{3}:|y|\leq r\}
\end{equation}
and $\exp _{p}:B_{E}(0,r)\rightarrow B(p,r)\subset \Sigma $ is a
diffeomorphism onto its image. The largest radius $r$ for which this is
true, as $p$ varies in $\Sigma $, is called the injectivity radius $\mathrm{
inj}_{M}$ of $(\Sigma ,g_{ab})$ (Fig.~\ref{fig:balldomains}).

\medskip

Let $B(p;r_{0})$ denote a given geodesic ball of radius $r_{0}<\mathrm{inj}
_{M}$, and for vector fields $X$, $Y$, and $Z$, in $(B(p;r),g_{ab})$ let $
\mathcal{R}(X,Y)Z=\mathcal{R}_{\;\,bcd}^{a}X^{c}Y^{d}Z^{b}$ $\partial _{a}$
be the corresponding curvature tensor. Since $r:(B(p;r),g_{ab})\rightarrow 
\mathbb{R}$ is a distance function (\emph{i.e.}, $|\nabla r|\equiv 1$), the
geometry of $B(p;r)$ can be described by the following set of equations \cite
{petersen}: 
\begin{eqnarray}
(\nabla _{\partial _{r}}S)(X)+S^{2}(X)=-{\cal R}(X,\partial _{r})\partial
_{r},  \label{shape} \\
({\cal L }_{\partial _{r}}g)(X,Y)=2g(S(X),Y),  \label{smetr} \\
\nabla _{\partial _{r}}S={\cal L }_{\partial _{r}}S,  \label{slie}
\end{eqnarray}
where $\partial _{r}=\nabla r$ is the gradient of $r$, ${\cal L }_{\partial
_{r}}g$ is the Lie derivative of the 3--metric $g$ in the radial direction $
\partial _{r}$, the shape tensor\footnote{
The Hessian $S_{ij}=\nabla _{ij}^{2}r$ is the second fundamental
form of the immersion $U_{r}\hookrightarrow (B(p;r),g_{ab})$. We use the
equivalent characterization of \emph{shape tensor} in order to avoid
confusion with the standard second fundamental form of use in relativity.}
$S=\nabla ^{2}r$ is the Hessian of $r$, and  $S^{2}(X)$ abbreviates 
$S^{2}(X)=S^{\alpha}_{\;\,\eta} S^{\eta}_{\;\,\beta} X^{\beta}\partial_{\alpha}$.
Such equations follow from the Gauss--Weingarten relations applied to
study the $r$--constant slices $U_{r}:=\{\exp _{p}(r\vec{v})\in \Sigma
:r=const.\}$, which are the images in $(B(p;r),g_{ab})$ of the standard
Euclidean $2$--spheres $\mathbb{S}^{2}(r)\subset T_{p}\Sigma \simeq \mathbb{R
}^{3}$. The shape tensor $S_{ab}$ measures how the bidimensional metric $
g^{(2)}(X,Y)$ induced on $U_{r}$ by the embedding in $(B(p;r),g_{ab})$
rescales as the radial distance $r$ varies. If we denote by $P_{U}(W):=
W-g(W,\partial _{r})\partial _{r}$ the tangential projection of a vector $
W\in T_{p}\Sigma $ onto the tangent space $T_{p}U$ to the surface $U_{r}$ at 
$p$, then together with (\ref{shape}), (\ref{smetr}), (\ref{slie}) we also
get the tangential curvature equation:
\begin{eqnarray}
P_{U}\left( {\cal R}(X,Y)Z\right)= \nonumber \\
\fl {\cal G}\left( g^{(2)}(Y,Z)X-g^{(2)}(X,Z)Y\right)
-g^{(2)}(S(Y),Z)S(X)-g^{(2)}(S(X),Z)S(Y)\;, 
\label{tang}
\end{eqnarray}
where the vectors $X$ and $Y$ are tangent to $U_{r}$, and where $\cal G$
denotes the Gaussian curvature of $(U_{r},g^{(2)})$.  Further properties of
the equations (\ref{shape}) and (\ref{smetr}) that we need are best seen by
using polar geodesic coordinates. Recall that
normal exponential coordinates at $p$ are geometrically defined by 
\begin{eqnarray}
\exp _{p}^{-1}:B(p;r)\rightarrow T_{p}\Sigma \simeq \mathbb{R}^{3} \;\;\;;\;\;\;
q\longmapsto \exp _{p}^{-1}(q)=(y^{i})\;\;,
\end{eqnarray}
where $(y^{i})$ are the Cartesian components of the velocity vector $\vec{v}
\in $ $T_{p}\Sigma $ characterizing the geodesic segment from $p$ to $q$.
Such coordinates are unique up to the chosen identification of $T_{p}\Sigma $
with $\mathbb{R}^{3}$. Since we are dealing with a radial rescaling, for our
purposes a suitable identification is the one associated with the use of
polar coordinates in $T_{p}\Sigma \simeq \mathbb{R}^{3}$. We therefore
introduce an orthonormal frame $\{e_{1},e_{2},e_{3}\}$ in $T_{p}\Sigma $ such
that $e_{1}:=\partial _{r}$ and with $\{e_{2},e_{3}\}$ an orthonormal frame
on the unit $2$--sphere $\mathbb{S}^{2}(1)\subset T_{p}\Sigma $. We can
extend such vector fields radially to the whole $T_{p}\Sigma $; we consider
also the dual coframe ${\theta ^{2},\theta ^{3}}$ associated with
${e_{2},e_{3}}$. The introduction of such a polar coordinate system in
$T_{p}\Sigma $ is independent of the metric $g_{ab}$ and thus is ideally
suited for discussing geometrically the $r$--scaling properties of $%
\left\langle \varrho \right\rangle _{B(p;r)}$ (see, however, Remark 2).
If we pull--back to $T_{p}\Sigma $ the metric $g$ of $B(p;r)\subset \Sigma
$, we get: 
\begin{equation}
g=dr^{2}+g(e_{\alpha },e_{\beta })\theta ^{\alpha }\theta ^{\beta },\;\alpha
,\beta =2,3,  \label{polar}
\end{equation}
where the components $g_{\alpha \beta }:=g(e_{\alpha },e_{\beta })$ $
=g^{(2)}(r,\vartheta ,\varphi )$ are functions of the polar coordinates $
(\vartheta ,\varphi )$ in $T_{p}\Sigma \simeq \mathbb{R}^{3}$, associated
with the coframe ${\theta ^{2},\theta ^{3}}$. Note that such a local
representation of the metric holds throughout the local chart $(B(p;r),\exp
_{p}^{-1})$ and not just at $p$. In Cartesian coordinates in $T_{p}\Sigma
\simeq \mathbb{R}^{3}$ one recovers the familiar expression: 
\begin{equation}
g_{ab}=\delta _{ab}-\frac{1}{3}\mathcal{R}_{akbl}(p)y^{k}y^{l}+\mathcal{O}%
\left( |y|^{3}\right) \;.  \label{cartesian}
\end{equation}
In polar geodesic coordinates we have 
$(x^{1} :=r, x^{2}:= \vartheta ,x^{3}:=\varphi )$ and $g_{\alpha 
\beta}^{(2)}=g_{\alpha \beta }$ with $2\leq \alpha,\beta \leq 3$. Thus the 
equations (\ref{shape}), and (\ref{smetr}) take the
explicit form: 
\begin{eqnarray}
\partial _{r}S_{\;\,\beta }^{\alpha }=-S_{\;\,\eta }^{\alpha }S_{\;\,\beta
}^{\eta }-{\cal R}_{\;\,r\beta r}^{\alpha }\;\;;\\
\partial _{r}g_{\alpha \beta }= 2 S_{\;\,\alpha }^{\eta }g_{\eta \beta }\;\;,
\label{peterspolar}
\end{eqnarray}
where ${\cal R}_{\;\,r\beta r}^{\alpha }$ denote the radial components
of the curvature tensor. According to the tangential Gauss--Codazzi
equation (\ref{tang}) we can also write 
\begin{equation}
R_{\;\,\mu \beta \nu }^{\alpha }={\cal G}\left( \delta _{\;\,\beta }^{\alpha
}g_{\nu \mu }-\delta_{\;\,\nu }^{\alpha }g_{\beta \mu }\right) -S_{\;\,\beta
}^{\alpha }S_{\nu \mu }+S_{\;\,\nu }^{\alpha }S_{\beta \mu }\;.
\end{equation}
If we introduce the tangential components of the Ricci tensor
according to  
\begin{equation}
{\cal R}_{\;\,\beta }^{\alpha }:= g^{kl}{\cal R}_{\;\,k\beta l}^{\alpha }={\cal
R}_{\;\,r\beta r}^{\alpha }+{\cal G}\delta _{\;\,\beta }^{\alpha
}-S_{\;\,\beta }^{\alpha }S+S_{\;\,\nu }^{\alpha }S_{\;\,\beta }^{\nu }\;,
\end{equation}
then we can rewrite the radial components of the curvature as  
\begin{equation}
{\cal R}_{\;\, r \beta r}^{\alpha } = {\cal R}_{\;\,\beta }^{\alpha }-
{\cal G}\delta _{\;\,\beta }^{\alpha }+ S_{\;\,\beta }^{\alpha } S -
S_{\;\,\nu}^{\alpha} S_{\;\,\beta }^{\nu }\;,
\end{equation}
and the first of equations (\ref{peterspolar}) becomes 
\begin{equation}
\partial _{r}S_{\;\,\beta }^{\alpha }= - {\cal R}_{\;\,\beta }^{\alpha }+
{\cal G}\delta _{\;\,\beta }^{\alpha } - S_{\;\,\beta }^{\alpha }S\;,
\label{Ricscal}
\end{equation}
where $S:=S_{\;\,\beta }^{\alpha }\delta _{\;\,\beta }^{\alpha }$  is the rate
of area expansion of $g^{(2)}(r,\vartheta ,\varphi )$. Note that by taking
the trace of both equations (\ref{peterspolar}) we get: 
\begin{eqnarray}
\label{factor}
\partial_{r}S= - S_{\;\,\nu }^{\alpha }S_{\;\,\alpha }^{\nu }- Ric(\partial
_{r},\partial _{r})\;,   \\
g^{\alpha \beta }\partial _{r}g_{\alpha \beta }=2S \;,
\end{eqnarray}
where $Ric(\partial _{r},\partial _{r}):=$ ${\cal R}_{\;\,r i r}^{i}$
denotes the $\partial _{r}$ component of the Ricci tensor, (the first
equation in (\ref{factor}) is nothing but the Jacobi operator coming from
the second variation of the area associated with $g^{(2)}(r,\vartheta
,\varphi )$). If we assume that the curvature ${\cal R}(X,Y)Z$ is given,
then, by fixing the $(\vartheta ,\varphi )$--dependence in the factorized
metric (\ref{polar}), we can consider (\ref{peterspolar}) as a system of
decoupled ordinary differential equations describing  the rescaling of the
geometry of $B(p;r)$ in terms of the one--parameter flow of  immersions $
\mathbb{S}^{2}(r)\mapsto $ $(U_{r},g^{(2)}(r,\vartheta ,\varphi ))$. Note in
particular that the shape tensor matrix $(S_{\;\,\beta }^{\alpha })$ is
characterized, from the equation (\ref{Ricscal}), as a functional of the
Ricci curvature of the ambient manifold $(\Sigma ,g_{ab})$.

\begin{figure}
\begin{center}
\includegraphics[width=7cm]{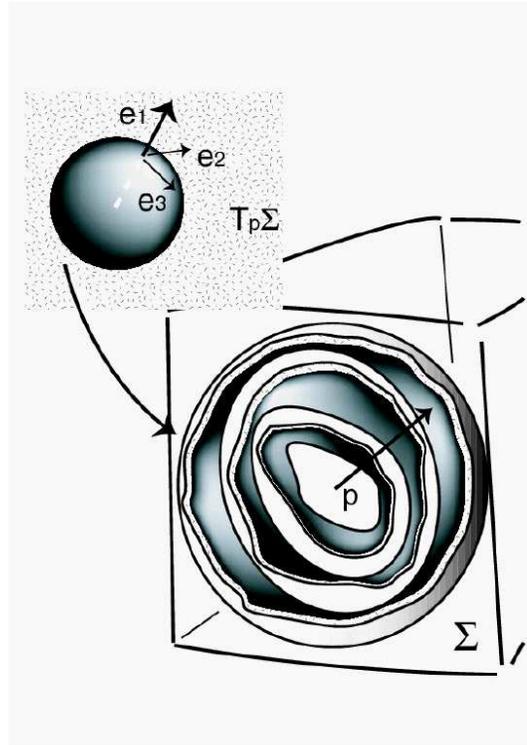}
\end{center}
\caption{\label{fig:petersen} 
A foliation of the geodesic ball along the radial direction into a family of geodesic 
balls is shown.
The metric on the boundary of the balls factorizes into radial and tangential components
due to our choice of the Dreibein $e_1,e_2,e_3$ in the ``master ball''.}
\end{figure}

\bigskip

{\sl Scaling properties of the metric}

\medskip

In such a setting 
the equation $\partial _{r}g_{\alpha \beta
}=2S_{\;\,\alpha }^{\eta }g_{\eta \beta }$ can be interpreted by saying
that the metric rescales radially along the (curvature--dependent) shape
tensor $S_{\;\,\beta }^{\alpha }$.  In order to get a more explicit
expression of such a radial rescaling, let us consider, at a fixed $r_{0}$, a
one--parameter family of surfaces $(U_{r},g^{(2)}(r,\vartheta ,\varphi ))$
with parameter $(r-r_{0})\geq 0$ starting at the given surface $
(U_{r_{0}},g^{(2)}(r_{0},\vartheta ,\varphi ))$ and foliating the ball $
B(p;r)$ (Fig.~\ref{fig:petersen}). 
In a sufficiently small neighborhood of the initial surface  $
(U_{r_{0}},g^{(2)}(r_{0},\vartheta ,\varphi ))$, we can write 
\begin{equation}
\fl S_{\;\,\beta }^{\alpha }(r)=S_{\;\,\beta }^{\alpha }(r_{0})+
\left[ - {\cal R}_{\;\,\beta }^{\alpha } + {\cal G}\delta^{\alpha}_{\;\,\beta} 
- S^{\alpha}_{\;\,\beta}S \right]
_{r_{0}}(r-r_{0})+\mathcal{O}((r-r_{0})^{2})\;.
\end{equation}
Inserting such an expression into $\partial _{r}g_{\alpha \beta
}=2S_{\;\,\alpha }^{\eta }g_{\eta \beta }$ we get
\begin{eqnarray}
\frac{1}{(r-r_{0})}\left[ \partial _{r}g_{\alpha \beta }(r)-\partial
_{r}g_{\alpha \beta }(r)\vert_{r_0}\right]  = \nonumber \\
-2{\cal R}_{\alpha \beta }(r_{0})+2\left[ {\cal G}\delta _{\alpha\beta}
-S_{\alpha\beta } S\right] _{r_{0}}+\mathcal{O}((r-r_0))\;. 
\label{radscale}
\end{eqnarray}
Note that the terms $2\left[ {\cal G}\delta _{\alpha\beta}-S_{\alpha\beta} S\right]_{r_{0}}$ 
only depend on the geometry (intrinsic and extrinsic) of the surface $(U_{r_{0}},g^{(2)}(r_{0},
\vartheta,\varphi ))$ and represent reaction tangential terms which work against the
curvature of the ambient $(\Sigma ,g_{ab})$. Thus, one may say that in a
sufficiently small neighborhood of the initial surface
$(U_{r_{0}},g^{(2)}(r_{0},\vartheta ,\varphi ))$ the ambient geometry of 
$(\Sigma ,g_{ab})$ forces the metric {\em to rescale radially in the
\emph{direction} of its Ricci tensor}. This latter remark will turn out quite
useful in understanding the geometric rationale behind the choice of a proper
averaging procedure for the geometry of $(\Sigma ,g_{ab})$.

\newpage

{\sl Scaling properties of the averaged density field}

\medskip

Guided by such geometrical features of geodesic balls, let us go back to
the study of the scaling properties of $\left\langle \varrho \right\rangle
_{B(p;r)}$. To this end, for any $r$ and $s$ such that $r+s< {\rm inj}_{\Sigma}$,
let us consider the one--parameter family of diffeomorphisms
\begin{eqnarray}
H_{s}:(\Sigma ,p)\rightarrow (\Sigma ,p) \nonumber \\
q=\exp _{p}[r_{q}(\partial _{r},e_{2},e_{3})]\longmapsto H_{q}(q):= \exp
_{p}[(r_{q}+s)(\partial _{r},e_{2},e_{3})] \;,
\end{eqnarray}
defined by flowing each point $q\in B(p;r)$ a distance $s$ along the
unique radial geodesic segment issued at $p\in \Sigma $ and passing 
through $q$. Let us remark that, for any $r$ such that $r_{0}\leq r< {\rm 
inj}_{M}$, we can formally write  
\begin{equation}
B(p;r)=H_{(r-r_{0})}B(p;r_{0}).
\end{equation}
Thus, in a neighborhood of $B(p;r_{0})$ and for sufficiently small $r$, we
get 
\begin{equation}
M(B(p;r)):= \int_{B(p;r)}\varrho \;d\mu
_{g}=\int_{B(p;r_{0})}H_{(r-r_{0})}^{\ast }(\varrho \;d\mu _{g}) , \label{pullM}
\end{equation}
where $H_{(r-r_{0})}^{\ast }(\varrho d\mu _{g})$ is the Riemannian measure
obtained by pulling back ($\varrho \,d\mu _{g}$) under the action of 
$H_{(r-r_{0})}$.
By differentiating (\ref{pullM}) with respect to $r$, we have: 
\begin{eqnarray}
\frac{d}{dr} M(B(p;r)) =\lim_{h\rightarrow 0}\frac{\left[ 
M(B(p;r+h))-M(B(p;r))
\right] }{h} \nonumber \\
=\lim_{h\rightarrow 0}\frac{\left[ \int_{B(p;r_{0})}H_{(r-r_{0})+h}^{\ast
}(\varrho \;d\mu _{g})-\int_{B(p;r_{0})}H_{(r-r_{0})}^{\ast }(\varrho \;d\mu _{g})
\right] }{h}\;\;.
\end{eqnarray}
Since $H_{(r-r_{0})+h}=H_{(r-r_{0})}\circ H_{h}$, we can write the above
expression as: 
\begin{eqnarray}
\lim_{h\rightarrow 0}\left[ \int_{B(p;r_{0})}\frac{H_{(r-r_{0})}^{\ast }
\left[ H_{h}^{\ast }(\varrho \;d\mu _{g})-(\varrho \;d\mu _{g})\right] }{h}\right] 
\nonumber \\
\fl =\lim_{h\rightarrow 0}\left[ \int_{B(p;r)}\frac{\left[ H_{h}^{\ast }(\varrho\;
d\mu _{g})-(\varrho \;d\mu _{g})\right] }{h}\right] \; 
=\; \int_{B(p;r)}\lim_{h\rightarrow 0}\frac{\left[ H_{h}^{\ast }(\varrho \;d\mu
_{g})-(\varrho \;d\mu _{g})\right] }{h}\;, 
\end{eqnarray}
from which it follows that (use ${\cal L}_{\partial_r} d\mu_g = {\rm div} (\partial_r 
) d\mu_g$
and ${\cal L}_{\partial_r}\varrho = d\varrho 
(\partial_r)=\frac{\partial}{\partial r}\varrho$): 
\begin{eqnarray}
\frac{d}{dr} M(B(p;r)) = \int_{B(p;r)} {\cal L}_{\partial_r} \varrho \;d\mu _{g}
\label{demme} \nonumber \\
\fl =\int_{B(p;r)}\left( {\cal L}_{\partial_r }\varrho +\varrho {\rm div} (\partial
_{r})\right) \;d\mu _{g} \;
=\; \int_{B(p;r)}\left( \frac{\partial}{\partial r} \varrho +\frac{1}{2} 
\varrho g^{ab}
\frac{\partial}{\partial r} g_{ab}\right)\; d\mu _{g}\;\;,
\end{eqnarray}
where ${\cal L}_{\partial_{r}}$ and ${\rm div} (\partial_{r})$ denote the Lie 
derivative along the vector field $\partial_{r}$ and its divergence,
respectively, and where we have exploited the well--known expression for the
Lie derivative of a volume density along $\partial_r$: 
\begin{equation}
\frac{\partial}{\partial 
r}\sqrt{g}=\frac{1}{2}\sqrt{g}g^{ab}\frac{\partial}{\partial r} g_{ab}\;\;.
\end{equation}

\newpage

\begin{figure}
\begin{center}
\includegraphics[width=7cm]{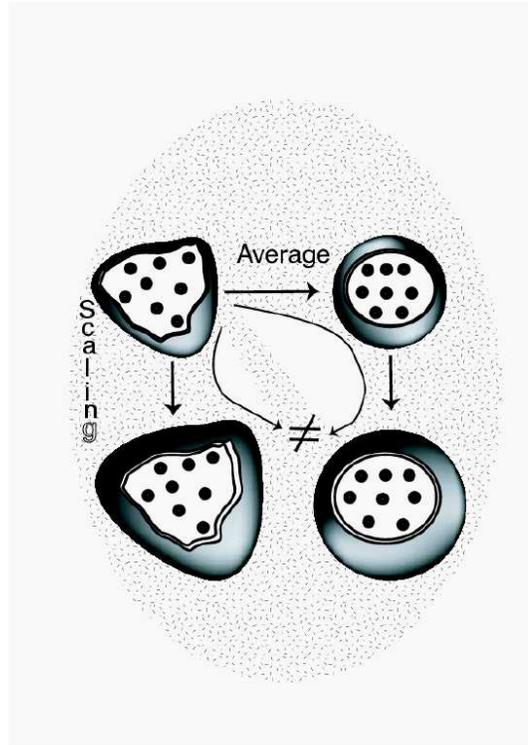}
\end{center}
\caption{\label{fig:noncommutativity} 
Volume--averaging scalar fields over the geodesic ball, and rescaling the radius of 
the ball are non--commuting operations. Note that a Lagrangian averaging as
well as a rescaling to first order in the radius preserves 
the material mass content of the ball.}
\end{figure}

With these preliminary remarks along the way, it is straightforward to
compute the total rate of variation with $r$ of the regional average $\left\langle
\varrho \right\rangle _{B(p;r)}$, since (\ref{demme}) implies: 
\begin{eqnarray}
\frac{\partial}{\partial r}\left\langle \varrho \right\rangle
_{B(p;r)}=\left\langle \frac{\partial}{\partial r} \varrho \right\rangle_{B(p;r)}\nonumber\\  
\fl +\;\frac{1}{2}\left\langle \varrho
g^{ab}\frac{\partial}{\partial r} g_{ab}\right\rangle
_{B(p;r)}\;-\;\frac{1}{2}\left\langle \varrho \right\rangle_{B(p;r)}\left\langle
g^{ab}\frac{\partial}{\partial r}g_{ab}\right\rangle_{B(p;r)}\;, 
\label{der}
\end{eqnarray}
where $\left\langle f\right\rangle _{B(p;r)}$ denotes the volume average of $
f$ over the ball $B(p;r)$. Explicitly, by exploiting (\ref{peterspolar}), we get 
\begin{equation}
\frac{\partial}{\partial r}\left\langle \varrho \right\rangle 
_{B(p;r)}=\left\langle 
\frac{\partial}{\partial r}
\varrho \right\rangle _{B(p;r)}+\left\langle \varrho
S\right\rangle _{B(p;r)}-\left\langle \varrho \right\rangle
_{B(p;r)}\left\langle S\right\rangle _{B(p;r)}.
\end{equation}
Thus, the regional average $\left\langle \varrho \right\rangle _{B(p;r)}$ feels
the fluctuations in the geometry as we vary the scale, fluctuations
represented by the shape tensor terms 
\begin{equation}
\left\langle \varrho S\right\rangle _{B(p;r)}-\left\langle \varrho 
\right\rangle
_{B(p;r)}\left\langle S\right\rangle _{B(p;r)}
\end{equation}
governed by the curvature in $B(p;r)$ according to (\ref{factor}), and
expressing a geometric non--commutativity between averaging over the ball 
$B(p;r)$ and rescaling its size (Fig.~\ref{fig:noncommutativity}). 
Since the curvature varies both in the given 
$B(p;r)$ and when we consider distinct base points $p$, the above remarks
indicate that the regional averages $\left\langle \varrho \right\rangle _{B(p;r)}$
are subjected to the accidents of the fluctuating geometry of $(\Sigma 
,g_{ab})$.
In other words, {\em there is no way of obtaining a proper smoothing of $\varrho$
without smoothing out at the same time the geometry of $(\Sigma ,g_{ab})$}.

\bigskip

{\sl Non--commutativity of averaging and scaling}

\medskip

It is straightforward to generalize the foregoing result for any 
smooth scalar function $\psi$ 
yielding a rule that summarizes the key result of this subsection.

\bigskip\noindent
{\bf Proposition 1} $\;\;$({\em Commutation Rule})

\smallskip\noindent 
{\em On a Riemannian hypersurface $(\Sigma, g_{ab})$
volume--averaging on a geodesic ball $B(p;r)$ and scaling (directional
derivative along the vector field $\frac{\partial}{\partial r}$) of a scalar
function $\psi$ are non--commuting operations,
as can be expressed by the rule:
\begin{eqnarray}
\frac{\partial}{\partial r}\langle\psi\rangle_{B(p;r)} - \langle 
\frac{\partial}{\partial r}\psi \rangle_{B(p;r)}=
\langle S \psi\rangle_{B(p;r)} - \langle S \rangle_{B(p;r)}
\langle\psi\rangle_{B(p;r)} \;;\nonumber \\  S  =
\frac{1}{\sqrt{g}} \frac{\partial}{\partial r}\sqrt{g}
\;\;;\;\; \langle S \rangle_{B(p;r)} = \frac{1}{V (B(p;r))}
\frac{\partial}{\partial r} V (B(p;r)) \;.
\label{commutation}  
\end{eqnarray} 

\medskip

Note that this formula may also be read as follows: 
\begin{equation}
\frac{\partial}{\partial r}\langle\psi\rangle_{B(p;r)} + 
\langle S \rangle_{B(p;r)}
\langle\psi\rangle_{B(p;r)} =  \langle \frac{\partial}{\partial r}\psi
+  S  \psi \rangle_{B(p;r)} \;.
\label{commutation2} 
\end{equation}
}

\newpage

\subsection{Eulerian averaging and Lagrangian smoothing}

The use of the exponential mapping in discussing the geometry behind the
regional averages $\left\langle \varrho \right\rangle _{B(p;r)}$ makes it clear
that we are trying to measure how different the averages $\left\langle \varrho
\right\rangle _{B(p;r)}$ are from the standard average over Euclidean balls.
In so doing we think of \ $\exp _{p}:T_{p}\Sigma \rightarrow \Sigma $ as
maps from the fixed space $B_{E}(0,r)$ into the manifold $(\Sigma ,g_{ab})$. In
this way we are implicitly trying to transfer information from the manifold $%
(\Sigma ,g_{ab})$ into domains of $\mathbb{R}^{3}$ which we would like to be, 
as far as possible, independent of the accidental geometry of $(\Sigma
,g_{ab})$ itself. Indeed, any averaging would be quite difficult to implement, 
if
the reference model varies with the geometry to be averaged. 
This latter task is only partially
accomplished by the exponential mapping, since the domain over which $\exp
_{p}:T_{p}\Sigma \rightarrow \Sigma $ is a diffeomorphism depends on $p$ and
on the actual geometry of $(\Sigma ,g_{ab})$. 
A suitable alternative is to use harmonic coordinates in the ball
$B(p;r)$, a technique which is briefly discussed in Remark 2.

We now go a step further by
considering not just the given $(\Sigma ,g_{ab})$, but rather a whole family of
Riemannian manifolds. To start with let us remark that, if we fix the
radius $r_{0}$ of the Euclidean ball $B_{E}(0;r_{0})\subset T_{p}\Sigma $
and consider the family of exponential mappings, $\exp _{(p,\beta
)}:T_{p}\Sigma \rightarrow (\Sigma ,g_{ab}(\beta ))$, associated with a
corresponding one--parameter family of Riemannian metrics $g_{ab}(\beta)$, 
$0\leq \beta <+\infty $, with $g_{ab}(\beta =0)=g_{ab}$, then $B(p;r_{0})$
becomes a functional of the set of Riemannian structures associated with $%
g_{ab}(\beta)$, $0\leq \beta <+\infty $,\emph{\ i.e.}, 
\begin{equation}
B(p;r_{0})\longmapsto B_{\beta }(p;r_{0}):= \exp _{(p,\beta )}\left[
B_{E}(0;r_{0})\right] .
\end{equation}

In this way, instead of considering just a given geodesic ball $B(p;r_{0})$,
we can consider, as $\beta $ varies, a family of geodesic balls  $%
B_{\beta }(p;r_{0})$, all with the same radius $r_{0}$ but with distinct
inner geometries $g_{ab}(\beta)$. Since $B_{\beta =0}(p;r_{0})=B(p;r_{0})$,
such balls can be thought of as being obtained from the given one $B(p;r_{0})$ by a
smooth continuous deformation of its original geometry. Under such
deformation also $\left\langle \varrho \right\rangle _{B(p;r_{0})}$ becomes 
$\beta-$dependent due to the functional dependence 
$\left\langle \varrho \right\rangle _{B_{\beta }(p;r_{0})}$.

\begin{figure}
\begin{center}
\includegraphics[width=13cm]{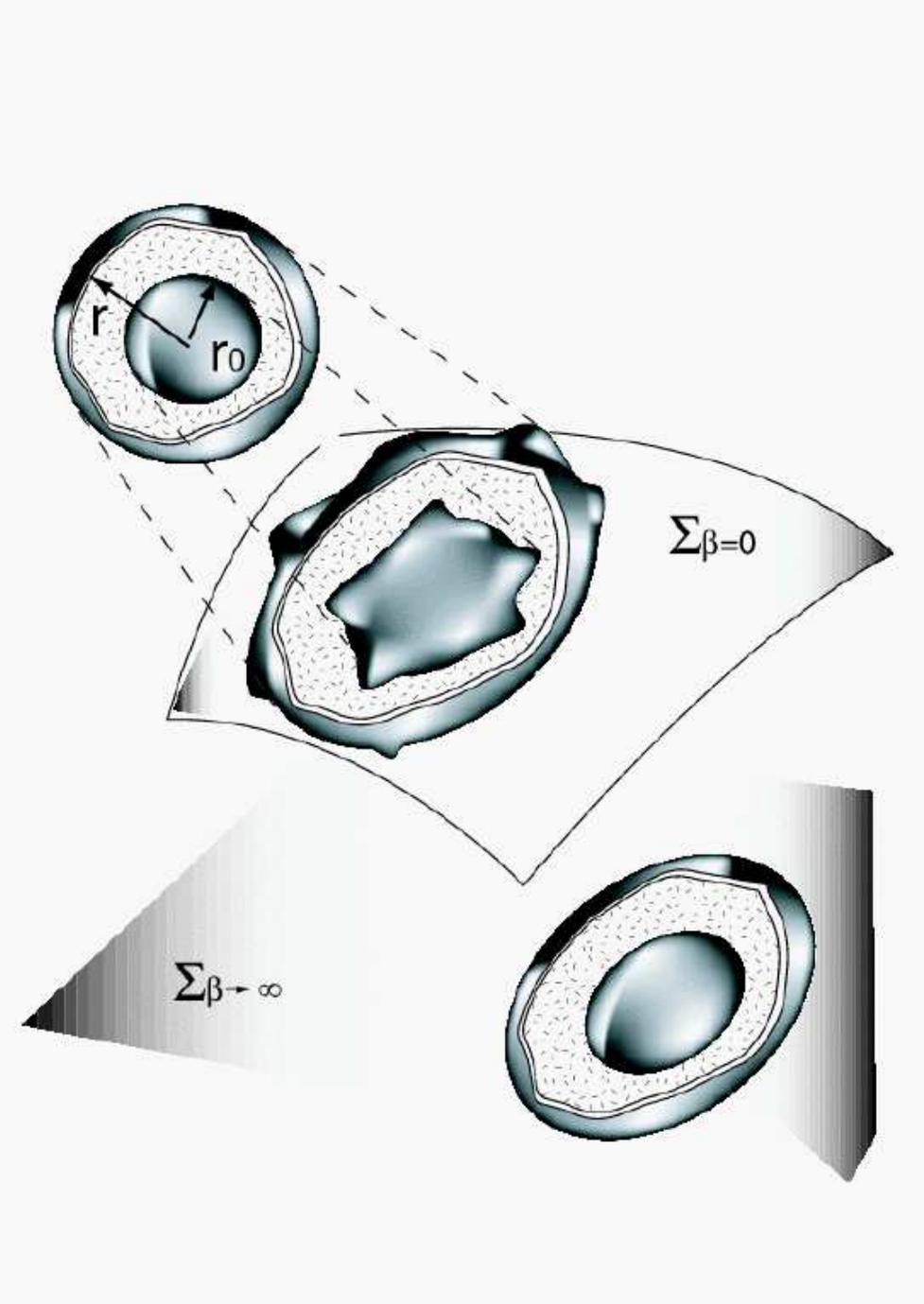}
\end{center}
\caption{\label{fig:averagingandscaling} 
Averaging over geodesic ball domains, while scaling the
geodesic radius is conceptually (and formally) equivalent with their 
Lagrangian deformation for fixed geodesic radius. The transformation
of a Lagrangian ball is illustrated with respect to $r-$scaling and 
$\beta-$deformation. 
}
\end{figure}

The elementary but basic observation in
order to take proper care of the geometrical fluctuations in 
$\left\langle \varrho \right\rangle_{B(p;r_{0})}$ is that the right member of
(\ref{der}) has precisely the formal structure of the linearization (\emph{
i.e.}, of the variation) of the functional $\left\langle \varrho \right\rangle
_{B(p;r_{0})}$ in the direction of the deformed
Riemannian metric $\frac{\partial}{\partial\beta} \lbrack
g_{ab}(\beta)\rbrack$,  \emph{viz.}, 
\begin{eqnarray}
\frac{\partial}{\partial\beta }\left\langle \varrho \right\rangle_{B_{\beta
}(p;r_{0})}=\left\langle \frac{\partial}{\partial\beta}\varrho \right\rangle
_{B_{\beta }(p;r_{0})}  \nonumber \\
\fl +\;\frac{1}{2}\left\langle \varrho g^{ab}(\beta )\frac{\partial }{\partial
\beta } g_{ab}(\beta )\right\rangle_{B_{\beta
}(p;r_{0})} \;-\; \frac{1}{2}\left\langle \varrho \right\rangle_{B_{\beta
}(p;r_{0})}\left\langle g^{ab}(\beta )\frac{\partial }{\partial \beta
}g_{ab}(\beta )\right\rangle _{B_{\beta }(p;r_{0})}\;,  
\label{geodef}
\end{eqnarray}
where the ball $B_{E}(0;r_{0})$ is \emph{kept fixed} while its image 
$B(p;r_{0})$ is deformed according to the flow of metrics $g_{ab}(\beta)$, 
$0\leq \beta \leq \infty$ (Fig.~\ref{fig:averagingandscaling}).

In a rather obvious sense, (\ref{geodef}) represents the \emph{active}
interpretation corresponding to the Eulerian \emph{passive} view associated
with the ball variation $B(p;r_{0})\rightarrow B(p;r)$. In other words, we
are here dealing with the (geometrical) Lagrangian point of view of following a
fluid domain in its deformation, where the \emph{fluid particles} here are the
points of $B(p;r_{0})$ suitably labelled. This latter remark suggests that
in order to optimize the averaging procedure associated with the regional
average $\left\langle \varrho \right\rangle _{B(p;r)}$, instead of studying its
scaling behavior as $r$ increases, and consequently be subjected to the
accidents of the fluctuating geometry of $(B(p;r),g_{ab})$, we may keep
fixed the domain $B_E(0;r_{0})$ (setting the scale over which we are
averaging) and rescale the geometry inside its image $B(p;r_{0})$ 
under the exponential map, according to a suitable flow of metrics $g_{ab}\rightarrow
g_{ab}(\beta)$, $0\leq \beta \leq \infty $. Correspondingly, also the
average matter density will be forced to rescale
$\langle\varrho\rangle_{B(p;r_{0})} \rightarrow 
\langle\varrho\rangle_{B_{\beta}(p;r_{0})} (\beta )$, and if we are able to
choose the flow $g_{ab}\rightarrow g_{ab}(\beta )$ in such a way that the
local inhomogeneities of the original geometry of $(\Sigma ,g_{ab})$ are
smoothly eliminated, then the regional averages $\left\langle \varrho
\right\rangle _{B_{\beta}(p;r_{0})}$ come closer and closer to represent a
matter averaging over a homogeneous geometry. 

\begin{figure}
\begin{center}
\includegraphics[width=11cm]{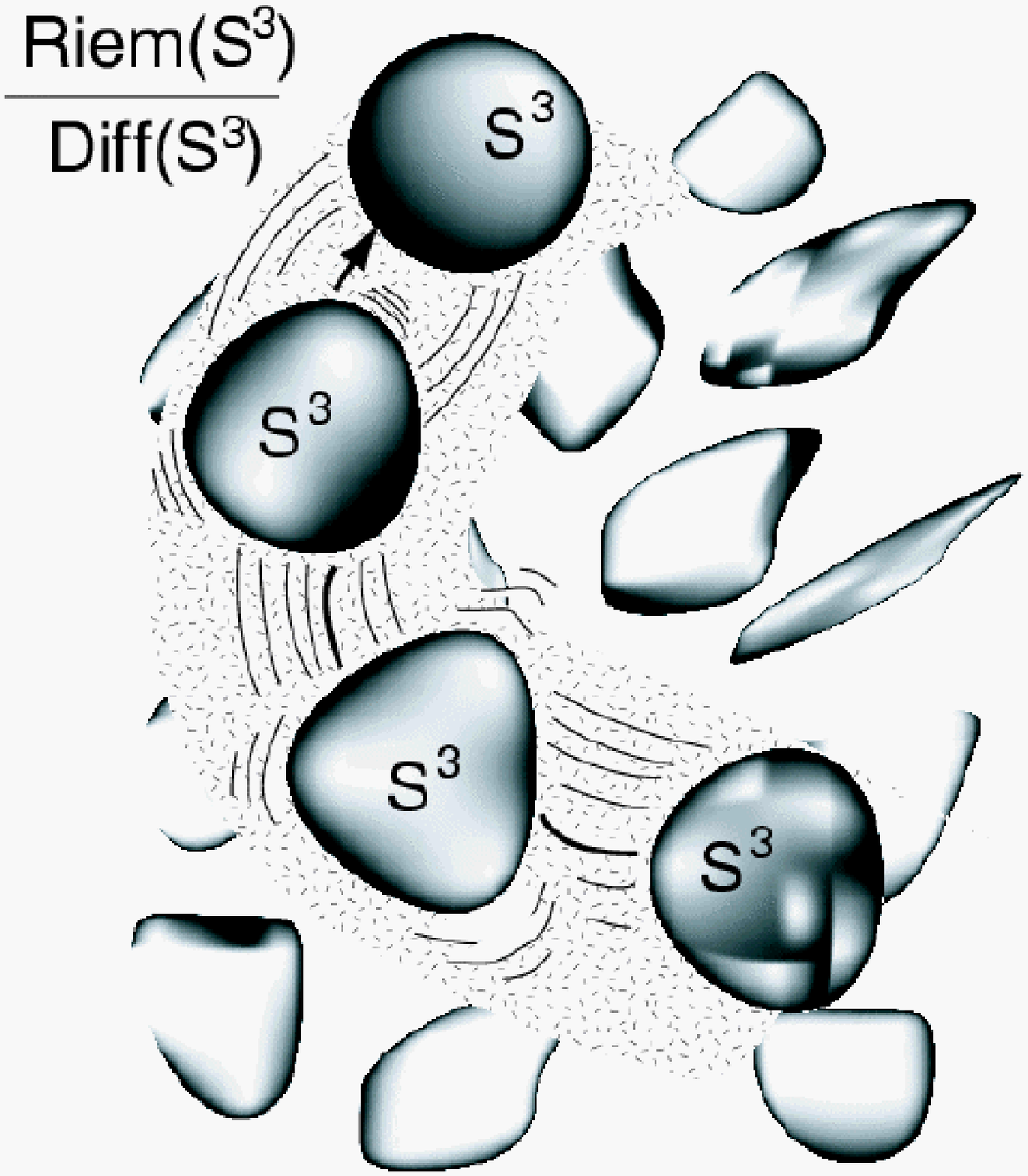}
\end{center}
\caption{\label{fig:ricciflow} 
When it is global, the Ricci flow acts in the space of all possible
Riemannian geometries by smoothing the geometry in a controllable fashion:
here we depict the case of a three--sphere whose maxima of the curvature
inhomogeneities are monotonically decreasing. If the original three--sphere
has positive Ricci curvature (i.e., is not wildly inhomogeneous), then it
undergoes a metamorphosis into the standard round three--sphere. Globally, if
large inhomogeneities are initially present, the Ricci flow may
experience singularities, the flow solution may feature a bifurcation
and consecutive regions may be 
`pinched off' in the limit where they are smoothed--out.
This implies a topology change of regional domains.
}
\end{figure}

According to the $r-$scaling properties of the metric
described by (\ref{radscale}), a natural candidate for such a Lagrangian flow
is the deformation generated by the Ricci tensor of the metric, 
a deformation flow that is strongly reminiscent of the Ricci flow on a 
Riemannian manifold $(\Sigma, g_{ab})$\cite{hamilton:ricciflow1}, \cite{carfora:deformation0},
\cite{hamilton:ricciflow2}: 
\begin{equation}
\frac{\partial }{\partial \eta }g_{ab}(\eta )=-2{\cal R}_{ab}(\eta )
\;\;\;;\;\;\; g_{ab}(\eta =0)=g_{ab}\;,
\label{riccifl}
\end{equation}
studied by Richard Hamilton and his co--workers in connection with an analytic
attempt to proving Thurston's geometrization conjecture. As is
well--known, the flow (\ref{riccifl}) is weakly--parabolic, and it is always
solvable for sufficiently small $\eta $. Obviously, it preserves any
symmetries of  $g_{ab}(\eta =0)=g_{ab}$. (The Ricci flow preserves the
isometry group of $(\Sigma, g_{ab})$.) It can be viewed as a ``heat
equation'' diffusing Riemannian curvature \cite{hamilton:ricciflow2} 
(Fig.~\ref{fig:ricciflow}).

\bigskip

{\sl Global normalization of the Ricci flow}

\medskip

The flow (\ref{riccifl}) may be reparametrized 
$\eta \rightarrow \beta $ by an $\eta$--dependent rescaling and by an 
$\eta$--dependent 
homothety $g_{ab}(\eta)\rightarrow \widetilde{g}_{ab}(\beta )$
so as to preserve the total volume of the manifold 
$(\Sigma,{\widetilde g}_{ab}(\beta))=:\Sigma_{\beta}$. 
For this end one has to introduce a suitably
normalized flow:
\begin{equation}
\fl \frac{\partial}{\partial \beta}\widetilde{g}_{ab}(\beta) = -2
\widetilde{\cal R}_{ab}(\beta)+ \widetilde{A}(\beta)\widetilde{g}_{ab}
(\beta)\;\;\;;\;\;\;\widetilde{g}_{ab}(\beta=0)=g_{ab}\;;
\label{normalizedflow}
\end{equation}
the normalization factor $\widetilde{A} (\beta)$ is determined in such a way 
that the volume of $\Sigma_{\beta}$ associated with the metric 
$\widetilde{g}_{ab}(\beta)$ does not change under deformation, 
$\partial / \partial\beta {\widetilde V}_{\Sigma_{\beta}}(\beta) = 0$. 
In place of (\ref{riccifl}) we then get the standard volume--preserving
Ricci flow that is usually studied in the mathematical literature
\cite{hamilton:ricciflow1},
\cite{hamilton:ricciflow2}:
\begin{equation}
\fl \frac{\partial}{\partial\lambda}\widetilde{g}_{ab}(\beta )=-2
\widetilde{\cal R}_{ab}(\beta )+\frac{2}{3}
\left\langle\widetilde{\cal R}(\beta)\right\rangle_{\Sigma_{\beta}}
\widetilde{g}_{ab}(\beta)
\;\;\;;\;\;\;\widetilde{g}_{ab} ( \beta = 0 ) = g_{ab} \;\;,
\label{unflow}
\end{equation}
whose global solutions (if attained) are constant curvature metrics:
\begin{equation}
\overline{g}_{ab}:=\lim_{\beta\rightarrow +\infty}\widetilde{g}_{ab}
(\beta)\;\;.
\end{equation}

\smallskip

Since the normalization factor
is spatially constant, it will not enter as a fluctuating quantity in our 
equations for the regional averages. If we would, e.g., normalize the Ricci flow
such that it preserves the global mass, this would not change the statements on
regional averages, where we shall require the preservation of the regional mass.
We emphasize that such a
normalization is a technical choice, mathematically needed in order to be
able to compare the distinct regional averagings carried out with respect to
balls with different centers.
Note that, even if we only wish to smooth the hypersurface $\Sigma _{\beta }$ 
on regions of (Euclidean) radius $r_{0}$, (i.e., $B_{E}(0;r_{0})$), 
their representatives $B_{\beta }(p;r_{0})$ are to be considered for distinct centers, 
say $p_{j}$; in other words, we can average over
$B_{\beta }(p_{1};r_{0})$, ..., $B_{\beta }(p_{k};r_{0})$, where $
\{p_{1},...,p_{k}\}$ is a set of points suitably scattered over the manifold
$\Sigma _{\beta }$ (compare Figure 3).
As a matter of fact, all our final results
factor out the global volume average and refer only to the average with
respect to the regional ball. It must also be stressed
that, to our knowledge, there is not yet a mathematically correct way for
implementing a Ricci flow that just works for an open region (such as a ball)
of a $3-$manifold (one needs suitable boundary conditions, on the spherical
boundary of the ball, controlling the flow of curvature in and/or out of
the ball). Since the total volume is not a physically
observable quantity, we take care of working out results which are
independent of the volume constraint.

\newpage


\section{Averaging and Scaling put into Practice}

\medskip

\subsection{Smoothing the metric}

Let us now come to the strategy for the optimal choice of the smoothing flow
for the metric $g_{ab}\rightarrow g_{ab}(\beta )$. As outlined in the previous section,
when dealing with regional averages, a suitably normalized Ricci flow taking
care of the metric comes naturally to the fore. 

\noindent 
To set notation,
let us again write the volume--preserving Ricci flow equations in the form:

\smallskip
\begin{equation}
\fl \frac{\partial }{\partial \beta }g_{ab}(\beta )=-2\mathcal{R}_{ab}(\beta )+
\frac{2}{3}g_{ab}(\beta )\langle \mathcal{R}(\beta )\rangle _{\Sigma _{\beta
}} \;\;\;;\;\;\;g_{ab}(\beta =0)=g_{ab}\;\;,
\label{mflow}
\end{equation}

\smallskip\noindent
where $0\leq \beta \leq +\infty $ is the deformation parameter. As already
stressed, on any compact Riemannian manifold $(\Sigma ,g_{ab})$ we have
Hamilton's theorem according to which a local solution to such a flow with $
g_{ab}(\beta =0)=g_{ab}$ always exists for $\beta$ sufficiently small
\cite{hamilton:ricciflow1}, \cite{deturck}; for
an introduction to such problematics, with many self--explanatory diagrams,
see \cite{carfora:deformation1}. The study of the existence and properties of 
global solutions $g_{ab}\rightarrow g_{ab}(\beta )$, $0\leq \beta <\infty $, 
is much more
difficult to establish, and is an active field of research (see the recent
review by R. Hamilton \cite{hamilton:ricciflow2}, and \cite{caochow}. 
In particular, if the initial metric $(\Sigma ,g_{ab})$ has positive
Ricci curvature, then the solution $g_{ab}\rightarrow g_{ab}(\beta )$ to 
(\ref{mflow}) exists for all $\beta $, and it converges exponentially fast,
as $\beta \rightarrow \infty $ to a constant positive sectional curvature
metric $(\Sigma ,\overline{g}_{ab})$, (forcing $\Sigma $ to be a space form
diffeomorphic to the $3-$sphere $\mathbb{S}^{3}$, possibly
quotiented by a finite group $\Gamma $ of isometries). Other examples of
a global Ricci flow are provided by those flows that evolve from locally
homogeneous metrics $(\Sigma ,g_{ab})$. In particular, the eight distinct
homogeneous geometries existing in dimension $n=3$ have been analyzed in
detail resulting in a non--singular global flow \cite{isenbergjackson};
for an example originating from relativity
see \cite{carfora:deformation2}. 

\begin{figure}
\begin{center}
\includegraphics[width=11cm]{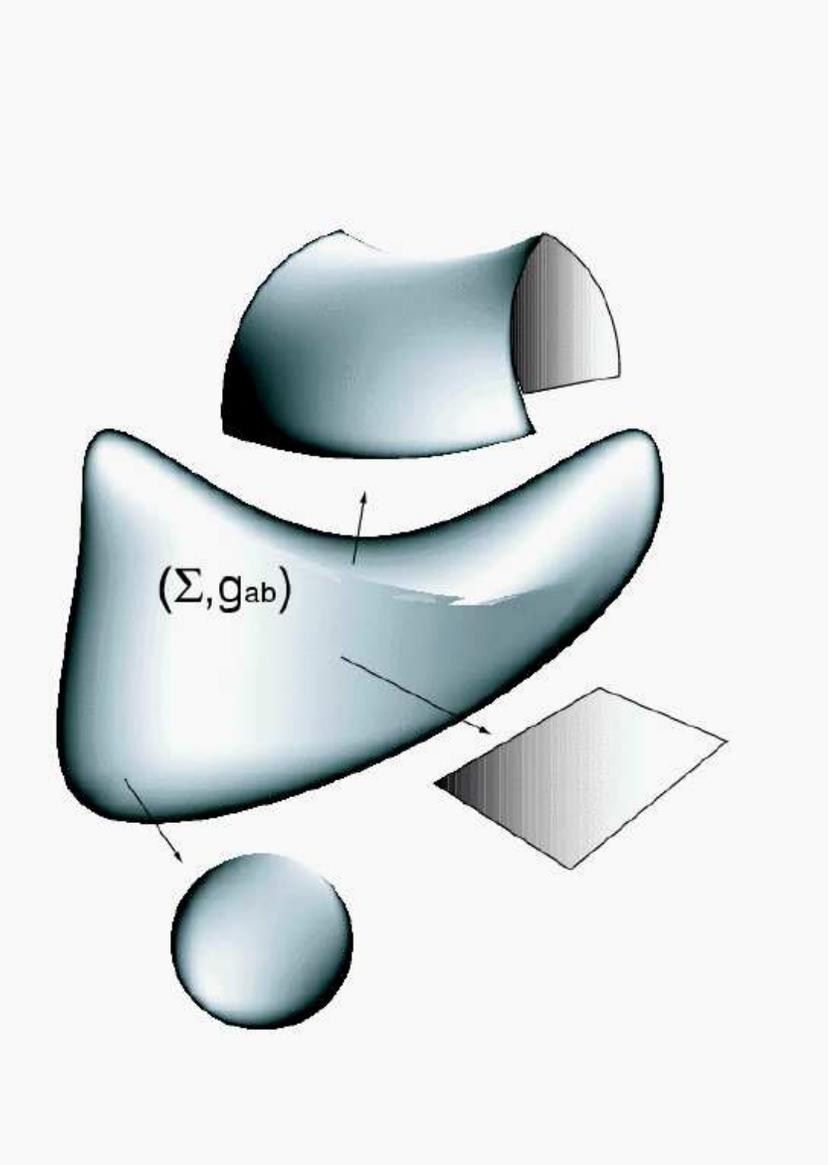}
\end{center}
\caption{\label{fig:thurstonconjecture} 
Thurston's conjecture as viewed by Richard Hamilton is illustrated. 
The global Ricci flow may have singularities resulting in 
an effective disconnection of the manifold into ``nice'' pieces. Thurston's
conjecture assumes that the original manifold can be cut into pieces which 
themselves have elementary Riemannian structures (associated with the eight 
``Thurston geometries''). The figure shows three such pieces in a two--dimensional
rendering which would correspond to pieces of the three possible FLRW space sections
in three dimensions:
the $3-$sphere, a piece of a $3-$cylinder, or a piece of a
$3-$hyperboloid.
}   
\end{figure}

\smallskip

Since, according to Thurston's geometrization conjecture 
(Fig.~\ref{fig:thurstonconjecture}),
every closed 3--manifold can be decomposed into pieces admitting one of the
eight geometric structures mentioned above, it is clear that the global
Ricci flow may play a distingushed role in such a conjecture. While such a
role clearly motivates the mathematical interest in (\ref{mflow}), it also
provides a strong argument in favour of (\ref{mflow}) as the natural
smoothing flow in a regional cosmological averaging procedure. 
As a matter of fact, the possibility of decomposing a 3--manifold into pieces
endowed with a locally homogeneous geometry is particularly appealing to
relativistic cosmology, where any such piece may be thought of as
representing the regional average of a sufficiently homogeneous portion of
the Universe. In such a framework it is 
suggestive to put the following result by Richard Hamilton 
into perspective \cite{hamilton:ricciflow3} (see also \cite{caochow}):

\newpage

\noindent
{\bf Theorem} $\;\;$({\em Hamilton})

\medskip
\noindent
If the closed 3--manifold $(\Sigma ,g_{ab})$ admits a non--singular solution $
g_{ab}\rightarrow g_{ab}(\beta )$ to (\ref{mflow}) for all $0\leq \beta
<\infty $, with uniformly bounded sectional curvature, then $(\Sigma ,
\overline{g}_{ab})$ can be decomposed into pieces admitting one of the
following locally homogeneous geometries:

\begin{itemize}

\item[(i)]{$(\Sigma ,\overline{g}_{ab})$ is a Seifert fibered space.}

\item[(ii)]{$(\Sigma ,\overline{g}_{ab})$ is a spherical space form $\mathbb{S}
^{3}/\Gamma $.}

\item[(iii)]{$(\Sigma ,\overline{g}_{ab})$ is a flat manifold.}

\item[(iv)]{$(\Sigma ,\overline{g}_{ab})$ is a constant negative sectional
curvature manifold.}

\item[(v)]{$(\Sigma ,\overline{g}_{ab})$ is the union (along incompressible tori)
of finite--volume constant negative sectional curvature manifolds and Seifert
fibered spaces.}

\end{itemize}

\medskip

The locally homogeneous geometries $(i)\cdots (v)$\ (in particular $(ii)$, $
(iii) $, and $(iv)$) are exactly the geometries after which, we believe,
the Universe can be regionally modelled after all accidental
inhomogeneities are \emph{ideally} ironed out. Thus, Hamilton's theorem
strongly advocates the basic role of the Ricci flow deformation as a
natural mean for averaging locally inhomogeneous 3--geometries in a
cosmological setting.

\bigskip

On the cosmological stage, however, we need also to discuss how such a
geometrical smoothing flow interacts with the actual distribution of matter.

\bigskip 

{\sl Combining the Ricci flow and the material mass flow}

\medskip

Our basic idea behind the regional averages $\left\langle \varrho \right\rangle
_{B(p;r)}$ is that they replace
the local accidental distribution of matter described by $\varrho $. Also,
instead of considering just a given geodesic ball $B(p;r_{0})$, we are
considering, as $\beta $ varies, a family of geodesic balls $B_{\beta
}(p;r_{0})$, all with the same radius $r_{0}$ but with distinct inner
geometries $g_{ab}(\beta )$. Note that $r_{0}$ plays here the role of a
distance cut--off: it is the typical scale over which we want to smooth the
empirical mass distribution. We \emph{have} to keep track of such a scale in
our setup. 

\begin{figure}
\begin{center}
\includegraphics[width=7cm]{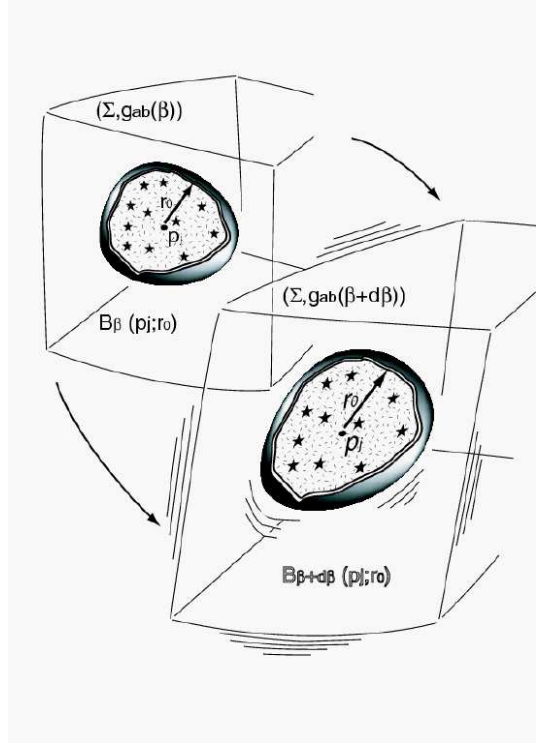}
\end{center}
\caption{\label{fig:riccimass} 
The conservation of the total material mass contained within geodesic ball domains 
guarantees the comparison of average properties on cosmological mass--scales. 
}   
\end{figure}

The key idea for obtaining a deformation equation for the regionally averaged
mass density field is to fix, besides the scale $r_{0}$ of the Lagrangian
ball, also the mass content of the ball under consideration during the
deformation process (Fig.~\ref{fig:riccimass}). 
It is a physically sensible idea
to concentrate on mass scales, since they are only directly related to so--called
``comoving'' length scales or volumes, respectively \cite{peebles}, if the 
Eulerian ball is also Euclidean and the inhomogeneites are set up relative to a global
reference flow providing the global comoving coordinates.

In order to obtain a deformation equation for the regionally averaged mass
density field in the chosen regions $B_{\beta }(p_{j};r_{0})$, we require
\begin{equation}
\frac{\partial }{\partial \beta }M\left( B_{\beta }(p_{j};r_{0})\right) =:
\frac{\partial }{\partial \beta }M_{\mathcal{B}_{\beta }}=0  \label{semiloc}
\end{equation}
along the solution of (\ref{mflow}). Since (the abbreviation $\mathcal{B}
_{\beta }:=B_{\beta }(p_{j};r_{0})$ is used hereafter)
\begin{equation}
\left\langle \varrho \right\rangle _{\mathcal{B}_{\beta }}(\beta )=\frac{M_{
\mathcal{B}_{\beta }}}{V_{\mathcal{B}_{\beta }}(\beta )}\;,
\label{averagedensity}
\end{equation}
where $V_{\mathcal{B}_{\beta }}$ denotes the (for each $\beta $ different)
volume of the ball in question, it is straightforward to check that we can
equivalently rewrite (\ref{semiloc}) as:
\begin{equation}
\frac{\partial }{\partial \beta }\ln 
\frac{\left\langle \varrho \right\rangle _{
\mathcal{B}_{\beta }}(\beta )}{\left\langle \varrho 
\right\rangle_{\mathcal{B}_0}}
+\frac{\partial }{\partial \beta }\ln \frac{V_{
\mathcal{B}_{\beta }}(\beta )}{V_{\mathcal{B}_0}}\;=\;0\;.
\end{equation}
We realize the local rate of volume deformation through the Ricci
deformation flow (\ref{mflow}) by exploiting the relation:
\begin{equation}
\frac{1}{\sqrt{g(\beta )}}\frac{\partial }{\partial \beta }\sqrt{g(\beta )}=
\frac{1}{2}g^{ab}(\beta )\frac{\partial }{\partial \beta }g_{ab}(\beta
)=\langle \mathcal{R}(\beta )\rangle _{\Sigma _{\beta }}-\mathcal{R}(\beta
)\;,  \label{Slocal}
\end{equation}
where $\mathcal{R}(\beta ):=g^{ab}(\beta )\mathcal{R}_{ab}(\beta )$ is the
scalar curvature associated with $g_{ab}(\beta )$. From this one easily
computes that
\begin{equation}
\frac{\partial }{\partial \beta }\ln 
\frac{V_{\mathcal{B}_{\beta }}(\beta)}{V_{\mathcal{B}_0}}\;=\;
-\left\langle\mathcal{R}(\beta )\right\rangle _{\mathcal{B}_{\beta }}+\left\langle
\mathcal{R}(\beta )\right\rangle _{\Sigma _{\beta }}\;.  \label{volev}
\end{equation}
Thus we get that the condition of the conservation of the matter content of $
B_{\beta }(p_{j};r_{0})$ (\ref{semiloc}) provides the following 
$\beta$--evolution for the regionally averaged mass density:
\begin{equation}
\frac{\partial }{\partial \beta }\ln \frac{\left\langle\varrho\right\rangle
_{\mathcal{B}_{\beta }}(\beta)}{\left\langle\varrho\right\rangle_{\mathcal{B}_0}}\;=\;
\left\langle \mathcal{R}(\beta )\right\rangle _{
\mathcal{B}_{\beta }}-\left\langle \mathcal{R}(\beta )\right\rangle _{\Sigma
_{\beta }}\;.  \label{rhoev}
\end{equation}

So far we have established an analogy between matter averaging on different
scales and geometrical deformation induced by a suitable flow of metrics.
With regard to the constraint equations of general relativity we have to
guarantee that this flow of metrics is compatible with the constraints.
Before we turn to the problem of smoothing the second fundamental form, it
is necessary but also illuminating in what follows to study stability properties
of the Ricci flow.

\subsection{Stability of the Ricci flow}

Associated with the Ricci flow we need to discuss also the properties of the
corresponding linearized flow. Roughly speaking, such a necessity comes
about since we may be asked what happens to our flow, if the original metric $
g_{ab}$, deformed according to $g_{ab}\rightarrow g_{ab}(\beta )$, is
slightly perturbed (which is a good question, since the flow is actually
perturbed in a dynamical situation, e.g., in the direction of the extrinsic
curvature tensor in time $\epsilon =t$):
\begin{equation}
g_{ab}\rightarrow \widehat{g}_{ab}=g_{ab}+\varepsilon h_{ab}\;\;,  \label{epsil}
\end{equation}
where $h_{ab}$ is a symmetric bilinear form, and $\epsilon $ a small
parameter. For notational convenience, let us set
\begin{equation}
\mathrm{RF}[g_{ab}(\beta )]:=-2\mathcal{R}_{ab}(\beta )+\frac{2}{3}
g_{ab}(\beta )\left\langle \mathcal{R}(\beta )\right\rangle _{\Sigma _{\beta
}}\;,
\end{equation}
so that the Ricci flow can be compactly written as
\begin{equation}
\frac{\partial }{\partial \beta }g_{ab}(\beta )=\mathrm{RF}[g_{ab}(\beta )]
\;\;\;;\;\;\;g_{ab}(\beta =0)=g_{ab}\;.
\end{equation}
It is easily checked that, if we perturb the initial metric $g_{ab}(\beta
=0)=g_{ab}$, and evolve the perturbed metric $\widehat{g}_{ab}$ according to
(\ref{mflow}), then we get
\begin{equation}
\widehat{g}_{ab}\rightarrow \widehat{g}_{ab}(\beta )=g_{ab}(\beta
)+\varepsilon h_{ab}(\beta )+\mathcal{O}(\varepsilon ^{2}),
\end{equation}
where $h_{ab}\rightarrow h_{ab}(\beta )$ is a linear flow solution of 
\begin{eqnarray}
\fl \frac{\partial }{\partial \beta }h_{ab}(\beta )=D\mathrm{RF}[g_{ab}(\beta
)]\circ h_{ab}(\beta ):=\frac{d}{d\epsilon }\mathrm{RF}[g_{ab}(\beta
)+\varepsilon h_{ab}(\beta )]\Big|_{\varepsilon =0} \nonumber \\
\fl h_{ab}(\beta =0)=h_{ab}\;\;.
\end{eqnarray}
Explicitly, we obtain (dropping the explicit $\beta $--dependence for
notational ease),
\begin{eqnarray}
 \label{linear}
\frac{\partial }{\partial \beta }h_{ab}=\frac{2}{3}\left\langle \mathcal{R}
\right\rangle _{\Sigma _{\beta }}h_{ab}  \nonumber \\
+\frac{2}{3}g_{ab}\left[ \frac{1}{2}\left\langle \mathcal{R}
h_{ik}g^{ik}\right\rangle _{\Sigma _{\beta }}-\frac{1}{2}\left\langle
\mathcal{R}\right\rangle _{\Sigma _{\beta }}\left\langle
h_{ik}g^{ik}\right\rangle _{\Sigma _{\beta }}-\left\langle \mathcal{R}
^{ab}h_{ab}\right\rangle _{\Sigma _{\beta }}\right]  \nonumber \\
-\Delta _{L}h_{ab}+2\left[ \delta ^{\ast }\left( \delta \left( h-\frac{1}{2}(
\mathrm{tr}h)g\right) \right) \right] _{ab}\;\;,
\end{eqnarray}
where
\begin{equation}
\Delta _{L}h_{ab}:=-\nabla ^{s}\nabla _{s}h_{ab}+\mathcal{R}_{as}h_{b}^{s}+
\mathcal{R}_{bs}h_{a}^{s}-2\mathcal{R}_{asbt}h^{st}
\end{equation}
is the Lichnerowicz--deRham Laplacian on bilinear forms, the operator $
\delta $ is (minus) the divergence, and $\delta ^{\ast }$ is its formal $
L^{2}$--adjoint ( \emph{i.e.}, $1/2$ the Lie derivative operator). All such
operators are considered with respect to the $\beta $--varying metric $
g_{ab}(\beta )$ of the unperturbed Ricci flow (\ref{mflow}). It is
clear from its explicit expression that (\ref{linear}) takes quite a simpler
form, if we restrict our attention to traceless perturbations $h_{ab}$, $
\emph{i.e.}$, if 
\begin{equation}
h_{ab}g^{ab}=0\;\;.
\end{equation}

\begin{figure}
\begin{center}
\includegraphics[width=13cm]{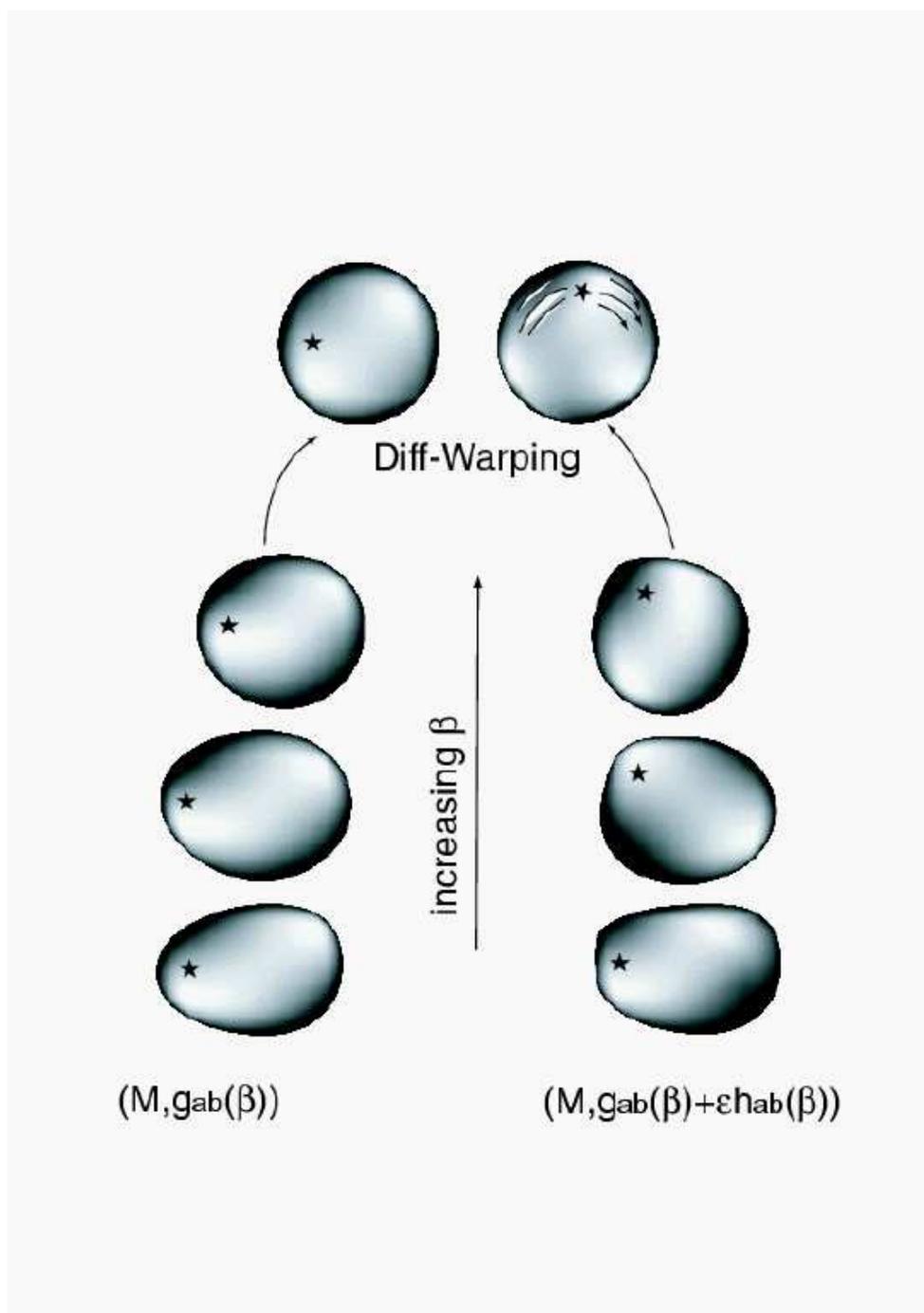}
\end{center}
\caption{\label{fig:stability} 
The stability analysis of the Ricci flow under symmetric perturbations allows us to 
determine the smoothing flow for the second fundamental form. It can be viewed to
represent an infinitesimal deformation of metrics connecting two neighboring 
flows of metrics. The flow of this deformation itself can be shown to be 
(weakly) parabolic (like the Ricci flow for the metric) up to the infinitesimal
equivariance of the Ricci flow under the diffeomorphisms group Diff($\Sigma$)
(``Diff--Warping'').   
}   
\end{figure}

It is then verified that such a condition, if it holds initially (for $\beta
=0$), it holds for each value of $\beta $ for which the flows (\ref{mflow})
and (\ref{linear}) are defined, and we get the much simpler result: 
\begin{equation}
\label{trless}
\fl \frac{\partial }{\partial \beta }h_{ab}=\frac{2}{3}\left\langle \mathcal{R}
\right\rangle _{\Sigma _{\beta }}h_{ab}-\frac{2}{3}g_{ab}\left\langle 
\mathcal{R}^{ab}h_{ab}\right\rangle _{\Sigma _{\beta }}  
-\Delta _{L}h_{ab}+{\pounds }_{-\nabla ^{k}h_{ik}}g_{ab}\;.
\end{equation}
For a given Ricci flow $g_{ab}\rightarrow g_{ab}(\beta )$, Eq.~(\ref{trless})
defines a linear (weakly) parabolic initial value problem (the strict
parabolicity is broken by the Lie derivative term\footnote{Note that 
$\pounds$ abbreviates the conformal (i.e., trace--free) Lie derivative, which should
not be confused with the Lie derivative denoted by $\cal L$.}:
\begin{equation}
\pounds _{-\nabla ^{k}h_{ik}}g_{ab}=-\nabla _{a}(\nabla ^{k}h_{kb})-\nabla
_{b}(\nabla ^{k}h_{ka})\;\;,
\end{equation}
associated to the infinitesimal equivariance under the diffeomorphisms group 
$\mathrm{Diff}(\Sigma )$). Given the initial (traceless) perturbation, $
h_{ab}(\beta =0)=h_{ab}$, the solution of (\ref{trless}) always exists and
is unique, and it represents an infinitesimal deformation of metrics
connecting two neighboring flows of metrics $g_{ab}\rightarrow g_{ab}(\beta
) $ and $\widehat{g}_{ab}\rightarrow \widehat{g}_{ab}(\beta )$. Since $
h_{ab}(\beta )g^{ab}(\beta )=0$, both flows have the same $\beta $
--dependent volume element,\emph{\ i.e.}, 
\begin{equation}
\sqrt{g(\beta )}=\sqrt{\widehat{g}(\beta )}\;\;,
\end{equation}
and thus the same $\beta $--dependent average density $\left\langle \varrho
\right\rangle _{\mathcal{B}_{\beta }}$. It is also important to remark that
the solution of (\ref{trless}) corresponding to the trivial initial datum (a
conformal Lie derivative term $\pounds _{\vec w}g_{ab}$), 
\begin{equation}
h_{ab}(\beta =0)=\pounds _{\vec w}g_{ab}:=\nabla _{a}w_{b}+\nabla _{b}w_{a}-\frac{%
2}{3}g_{ab}\nabla ^{c}w_{c}\;\;,
\end{equation}
where $w_{b}$ is a smooth ($\beta $--independent) vector field on $\Sigma $,
is provided by 
\begin{equation}
\fl h_{ab}(\beta )=\pounds _{\vec w}g_{ab}(\beta ):=\nabla (\beta )_{a}w_{b}+\nabla
(\beta )_{b}w_{a}-\frac{2}{3}g_{ab}(\beta )\nabla (\beta )^{c}w_{c}\;\;,
\end{equation}
where the $\beta $--dependence is only through the flow of metrics $%
g_{ab}\rightarrow g_{ab}(\beta )$ and the associated connection $\nabla
_{a}(\beta )$. Such a property is simply a consequence of the $\mathrm{Diff}%
(\Sigma )$ equivariance of the Ricci flow (Fig.~\ref{fig:stability}). 
By exploiting such a result, it
is possible to prove an important factorization theorem \cite{lott} for the
structure of the solution $h_{ab}\rightarrow h_{ab}(\beta )$ of (\ref{trless}),
which will prove invaluable throughout the rest of the paper, \emph{viz.},

\newpage

\noindent
{\bf Proposition 2} $\;\;$({\em Lott})

\smallskip\noindent
{\em If $h_{ab}\rightarrow h_{ab}(\beta )$ is the flow solution of (\ref
{trless}) corresponding to the initial (traceless) datum $h_{ab}(\beta
=0)=h_{ab}$, then it can always be factorized according to 
\begin{equation}
h_{ab}(\beta )=h_{ab}^{\ast }(\beta )+\pounds _{{\vec v}(\beta )}g_{ab}(\beta )
\;\;,
\end{equation}
where the bilinear form is the solution of the parabolic initial value
problem 
\begin{equation}
\frac{\partial }{\partial \beta }h_{ab}^{\ast }=-\Delta _{L}h_{ab}^{\ast }+
\frac{2}{3}\left\langle \mathcal{R}\right\rangle _{_{\Sigma _{\beta
}}}h_{ab}^{\ast }-\frac{2}{3}g_{ab}\left\langle \mathcal{R}^{ab}h_{ab}^{\ast
}\right\rangle _{_{\Sigma (\beta )}},
\end{equation}
with $h_{ab}^{\ast }(\beta =0)=h_{ab}$, and where the (now $\beta -$
dependent) vector field $v_{a}(\beta )$ is the flow solution of 
\begin{equation}
\frac{\partial }{\partial \beta }v_{a}(\beta )=-\nabla ^{c}h_{ac}^{\ast
}(\beta )\;\;,\;\;v_{a}(\beta =0)=0\;.
\end{equation}
}
\medskip

It is appropriate at this point to recall a few relevant facts concerning
the geometry behind the structure of the solutions of (\ref{trless}). It
follows from the above proposition that, as $\beta \rightarrow \infty $, $
h_{ab}(\beta )$ may either approach a (conformal) Lie derivative term $
\pounds _{\vec v}g_{ab}$, or a non--vanishing deformation tensor $h_{ab}^{\ast }$.
This latter non--trivial deformation is only present, if the corresponding
Ricci flow $g_{ab}\rightarrow g_{ab}(\beta )$ approaches an Einstein metric
on $\Sigma $ which is not isolated (for instance flat tori). In such a
case, there is a finite--dimensional manifold of such Einstein metrics, and
the non--trivial $h_{ab}^{\ast }$ simply represents infinitesimal
deformations connecting two infinitesimally neighboring Einstein metrics on $
\Sigma $. As is known, the round metric $\overline{g} _{ab}$ on the
three--sphere $\mathbb{S}^{3}$ is isolated in the sense that there are not
volume--preserving infinitesimal deformations of $\overline{g} _{ab}$
mapping it to another inequivalent constant curvature metric $\overline{g}
_{ab}^{^{\prime }}$. In this latter case, (\emph{i.e.}, for isolated
constant curvature metrics), as $\beta \rightarrow \infty $, $h_{ab}(\beta )$
must necessarily approach a (conformal) Lie derivative term $\pounds _{\vec v}
\overline{g}_{ab}$.

\subsection{Smoothing the second fundamental form}

The properties of the linearized Ricci flow for a traceless metric
perturbation $h_{ab}(\beta )$ naturally put to the fore an explicit way for
averaging the part of initial data $(\Sigma ,g_{ab},K_{ab},\varrho ,J_{a})$
related to the second fundamental form $K_{ab}$. One may contend that since $%
K_{ab}$ carries information on the way $(\Sigma ,g_{ab})$ is embedded in the
spacetime $(M^{(4)}\simeq \Sigma \times \mathbb{R},g^{(4)})$, one should
devise some way of deforming $K_{ab}$ which is independent of the flow of
metrics, since this latter flow only depends on the intrinsic geometry of $%
(\Sigma ,g_{ab})$. However, the very geometrical meaning of $K_{ab}$ shows
that such a point of view is not correct. According to the evolutive part of
the Einstein equations we have: 
\begin{equation}
K_{ab}=-\frac{1}{2N}\frac{\partial g_{ab}}{\partial t}+\frac{1}{2N}
{\cal L}_{\vec{N}}g_{ab}\;.
\end{equation}
Thus, we can write 
\begin{equation}
g_{ab}(t)=g_{ab}-t\left[ 2NK_{ab}-{\cal L}_{\vec{N}}g_{ab}\right] +\mathcal{O}%
(t^{2})\;\;,  \label{defmean}
\end{equation}
which clearly shows that $2NK_{ab}$ has the natural meaning of the
deformation tensor connecting two neighboring Riemannian metrics.
If the Ricci flow is the chosen averaging
procedure for deforming the metric in the initial data set $(\Sigma
,g_{ab},K_{ab},\varrho ,J_{a})$, then the stability (under small
perturbations) of such a deformation procedure requires that $NK_{ab}$ \emph{
must necessarily be} deformed according to the linearized flow (\ref{trless}
); \emph{there are no other consistent choices}.

The only freedom we have concerns which (algebraically independent) part of $
K_{ab}$ we want to deform according to (\ref{trless}). From the properties
of this latter flow it follows that a smart choice would be to leave
undeformed the trace part of $K_{ab}$, and deform only its traceless part
(i.e., the associated shear tensor $\sigma _{ab}:= K_{ab}-\frac{1}{3}g_{ab}K$).

\begin{figure}
\begin{center}
\includegraphics[width=12cm]{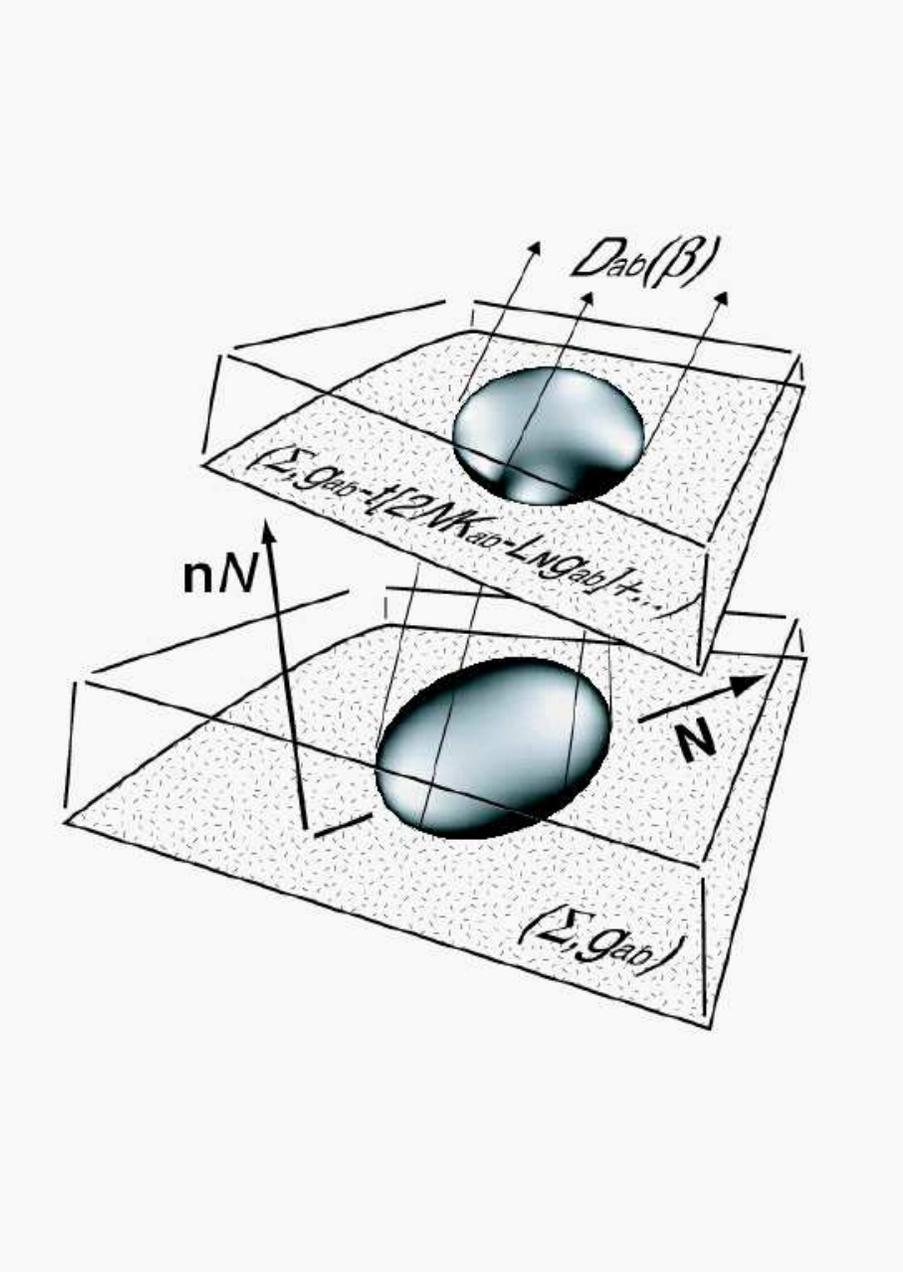}
\end{center}
\caption{\label{fig:distortiontensor} 
Smoothing the second fundamental form we have the freedom to choose which of the  
algebraically independent parts of the second fundamental form we wish to smooth out.
The trace--free distortion tensor is the natural choice in view of 
the evolutive part of Einstein's equations. In the embedding shown in the figure
we have assigned the coordinate time to the perturbation parameter $\varepsilon$, and the
perturbing symmetric bilinear form was associated with the second fundamental form.   
}   
\end{figure}

Actually, the geometrical meaning of $K_{ab}$ epitomized by (\ref{defmean})
suggests as a more natural choice that we deform the trace--free distortion
tensor (Fig.~\ref{fig:distortiontensor}): 
\begin{eqnarray}
\mathcal{D}_{ab} :=2N\sigma _{ab}-{\pounds }_{\vec{N}}g_{ab} \;\;\; \nonumber \\
\;\;\;\;\;\;=2N\sigma _{ab}\;-\left( \nabla _{a}N_{b}+\nabla _{b}N_{a}-\frac{2}{3}
g_{ab}\nabla ^{c}N_{c}\right) \;.
\label{distortiontensor}
\end{eqnarray}
In order to $\beta $--deform $\mathcal{D}_{ab}$ in a way consistent with the
Ricci flow, let us start observing that, with respect to the metric $
g_{ab} $, we can always decompose the given $\mathcal{D}_{ab}$ according to 
\begin{equation}
\mathcal{D}_{ab}=\mathcal{D}_{\perp ab}+\mathcal{D}_{\shortparallel ab}
\end{equation}
where $\mathcal{D}_{\perp ab}$ is the divergence--free part, 
\begin{equation}
\nabla ^{a}\mathcal{D}_{\perp ab}=0\;,
\end{equation}
and $\mathcal{D}_{\shortparallel ab}$ is the longitudinal part, 
\begin{equation}
\mathcal{D}_{\shortparallel ab}={\pounds }_{\vec w} g_{ab}\;,
\end{equation}
generated by the vector field $\vec w$ as a solution of the
(elliptic) partial differential equation:
\begin{equation}
\nabla ^{a}({\pounds }_{\vec w} g_{ab})=\nabla ^{a}\left( \nabla
_{a}w_{b}+\nabla _{b}w_{a}-\frac{2}{3}g_{ab}\nabla ^{c}w_{c}\right) =\nabla
^{a}\mathcal{D}_{ab}\;.
\end{equation}
Observe further that, according to (\ref{defmean}), $\mathcal{D}_{\perp ab}$
represents the part of the distortion tensor that deforms $g_{ab}$ into a
nearby distinct Riemannian structure $[g_{ab}(t)]=g_{ab}-\mathcal{D}_{\perp
ab}\,t+\mathcal{O}(t^{2})$, whereas $\mathcal{D}_{\shortparallel ab}$ simply
generates an infinitesimal ${\rm Diff}(\Sigma )$ reparametrization $
g_{ab}\rightarrow g_{ab}-t\left[ {\pounds }_{\vec w} g_{ab}-{\cal L}_{
\vec{N}}g_{ab}\right] +\mathcal{O}(t^{2})$.

With all this in mind, we can apply Lott's factorization theorem and
generate a natural smoothing flow for $\mathcal{D}_{ab}$, 
$\beta \rightarrow \mathcal{D}_{ab}(\beta )$ as a solution of 
\begin{eqnarray}
\label{alliscio}
\frac{\partial }{\partial \beta }\mathcal{D}_{ab}(\beta ) &=&\frac{2}{3}
\left\langle \mathcal{R}\right\rangle _{\Sigma _{\beta }}\mathcal{D}
_{ab}(\beta )-\frac{2}{3}g_{ab}\left\langle \mathcal{R}^{ab}\mathcal{D}
_{ab}(\beta )\right\rangle _{\Sigma _{\beta }}  \nonumber \\
&&-\Delta _{L}\mathcal{D}_{ab}(\beta )+{\pounds }_{-\nabla ^{k}\mathcal{D}
_{ik}(\beta )}g_{ab}\;.
\end{eqnarray}
by setting $\mathcal{D}_{ab}(\beta )=\mathcal{D}_{ab}^{\ast }(\beta )+$\ ${
\pounds }_{\vec v}g_{ab}(\beta )$, where $\mathcal{D}_{ab}^{\ast }(\beta )$
evolves according to the parabolic partial differential equation
\begin{equation}
\fl \frac{\partial }{\partial \beta }\mathcal{D}_{ab}^{\ast }(\beta )=-\Delta
_{L}\mathcal{D}_{ab}^{\ast }(\beta )+\frac{2}{3}\left\langle \mathcal{R}
\right\rangle _{_{\Sigma _{\beta }}}\mathcal{D}_{ab}^{\ast }(\beta )-\frac{2
}{3}g_{ab}\left\langle \mathcal{R}^{ab}\mathcal{D}_{ab}^{\ast }(\beta
)\right\rangle _{_{\Sigma (\beta )}}\;, 
\label{par1}
\end{equation}
with $\mathcal{D}_{ab}^{\ast }(\beta =0)=\mathcal{D}_{ab}$, and where the $
\beta -$dependent vector field $v_{a}(\beta )$ is the flow solution of 
\begin{equation}
\frac{\partial }{\partial \beta }v_{a}(\beta )=-\nabla ^{c}\mathcal{D}
_{ac}^{\ast }(\beta )\;\;,\;\;v_{a}(\beta =0)=0\;.  \label{par2}
\end{equation}
Corresponding to a global solution $\beta \rightarrow g_{ab}(\beta )$ of the
Ricci flow, that evolves towards an isolated constant curvature metric $\overline{
g}_{ab}=\lim_{\beta \rightarrow +\infty }g_{ab}(\beta )$, the (unique)
solution $\beta \rightarrow (\mathcal{D}_{ab}^{\ast }(\beta ),v_{a}(\beta ))$
of (\ref{par1}), (\ref{par2}) is such that $\overline{\mathcal{D}}_{\perp
ab}=\lim_{\beta \rightarrow +\infty }\mathcal{D}_{\perp ab}(\beta )=0$, ($
\overline{\mathcal{D}}_{\perp ab}$ will be different from zero only if the
constant curvature metric $\overline{g}_{ab}$ is not isolated), and $
\overline{\mathcal{D}}_{ab}$ reduces to a pure longitudinal shear,\emph{\
i.e.}, 
\begin{equation}
\overline{\mathcal{D}}_{ab}=\lim_{\beta \rightarrow +\infty }\mathcal{D}
_{ab}(\beta )={\pounds }_{\vec{\widehat{v}}}\overline{g}_{ab}\;,
\label{leshear}
\end{equation}
where 
\begin{equation}
\;\widehat{v}_{a}=\lim_{\beta \rightarrow +\infty }v_{a}(\beta )\;\;.
\label{finshift}
\end{equation}
Note that since we can always add to $\mathcal{D}_{ab}(\beta )$ the
trivial solution ${\pounds }_{\vec w}g_{ab}(\beta )$ of (\ref
{alliscio}), (where the vector field $\vec w$ does not depend
on $\beta $), by suitably choosing the ($\beta $--independent) vector field $
{\vec w}$, we can always assume that the conformal Lie derivative
term ${\pounds }_{\;\vec {\widehat{v}}}\overline{g}_{ab}$ provided
by (\ref{leshear}) comprises all the longitudinal shear present in $
\overline{\mathcal{D}}_{ab}$.

\subsection{The choice of a smoothing reference frame}

It is important to stress that the averaging of the trace--free distortion
tensor described above implies that we have made a consistent selection for
the lapse function $N$, the shift vector field $N_{b}$, and the rate of
volume expansion $K$ associated with the given initial data set $(\Sigma
,g_{ab},K_{ab},\varrho ,J_{a})$. As is well--known, such $(N,N_{b},K)$ are
kinematical quantities pertaining to the choice of the initial slice $
(\Sigma ,g_{ab},\mathcal{D}_{ab})$. They are related also to the choice
of the foliation $\Sigma _{t}$ in a suitably small (time) neighborhood of $
\Sigma $ in the spacetime resulting from the time--evolution of the data $
(\Sigma ,g_{ab},K_{ab},\varrho ,J_{a})$. Explicity, we have
\begin{equation}
\frac{\partial }{\partial t}K=-\Delta _{g}N+N\left[ R+K^{2}-3(4\pi G\rho
+\Lambda )\right] +N^{a}\nabla _{a}K\;\;.  \label{raych}
\end{equation}
This is basically Raychaudhuri's equation. The rationale underlying
the choice of a proper reference frame for carrying out a sensible
(regional) averaging, is that (\ref{raych}) should reflect the achieved
regularity in the averaged geometry and matter fields, without introducing
frame--dependent inhomogeneities and anisotropies. This implies that, as
we $\beta $--deform the local inhomogeneities and anisotropies of $(\Sigma
,g_{ab},\mathcal{D}_{ab})$, we need also to eliminate the inhomogeneous
artifacts due to the choice of the slicing associated with $(\Sigma
,N,N_{b},K)$. To give an example, it is often argued
that a good candidate for supporting an ``almost--Friedmannian''
initial data set $(\Sigma ,g_{ab},K_{ab})$ is the surface of constant matter
density in the cosmic fluid frame, or the surface of constant 4--velocity
potential (\cite{ellisstoeger}, \cite{ellisbruni}, \cite{bruni92a}, 
\cite{bruni92b}, \cite{dunsby}, \cite{carfora:RG}, \cite{buchert:grgfluid}). 
The point is that in such a slicing a slightly perturbed model features an
almost constant lapse function, since the instantaneous
acceleration for such a frame of reference is observationally quite small.
The observed expansion and shear are also small. In line of principle
these remarks suggest that a set $(\Sigma ,N,N_{b},K)$ characterized in this way
is, in the observed domain, the most suitable one for implementing an
averaging procedure. At any rate, even for such a natural and almost optimal
choice of $(\Sigma ,N,N_{b},K)$, we still have the issue of how to
consistently get rid of the (residual) frame fluctuations present in $N$, $
N_{b}$, and $K$. In order to eliminate such fluctuations, both the lapse
function $N(\beta )$, the shift vector field $N_{a}(\beta )$, and the rate
of volume expansion $K(\beta )$ are to be considered as explicitly 
$\beta-$dependent, (taking the given values $(N,N_{b},K)$ for $\beta =0$). 
In other words, we need to consider on $(\Sigma ,g_{ab},\mathcal{D}_{ab})$ a
one--parameter ($\beta $) family of (instantaneous) frames of reference $
(\Sigma ,N(\beta ),N_{b}(\beta ))$ and devise a way for characterizing their 
$\beta $--evolution. The rationale is to end up in a frame which, for $
\beta \rightarrow \infty $, reflects, as much as possible, the homogeneous
and isotropic properties of the geometry resulting from the $\beta $
--evolution of $(\Sigma ,g_{ab},\mathcal{D}_{ab})$. Such a frame will
correspond to the standard Friedmannian scenario of use in cosmology.

A word of caution is mandatory here. In mathematical cosmology it is often the
case that the choice of foliation is strictly connected to the structure of
the constraint equations via the use of $K$ as the variable
conjugated to time (York's extrinsic time). Spatially constant $K$ is then a
rather popular choice. However, the rate of volume expansion plays a
distinguished dynamical role in Friedmannian cosmology, and for
our purposes it would be quite detrimental to use $K$ as the variable
selecting the hypersurface $\Sigma $ carrying the data to be averaged. In
line of principle, the structure of the Ricci flow suggests that one should
pick up a $\Sigma $ such that $(\Sigma ,g_{ab})$ admits a global $0\leq
\beta <\infty $ Ricci flow, for instance this is the case if the Ricci
tensor of  $(\Sigma ,g_{ab})$ is positive or (hopefully!) not too wildly
oscillating. However, this is again something which is under 
control of mathematics, but not acceptable for the physical situation;
it cannot be used as a viable selection criterium. 
The best we can hope for is to assume that $\Sigma $
is ideally chosen among the $t$--constant slices of a global frame of
reference where, for scales sufficiently larger than the relevant averaging
scale, the distant galaxies appear to recede radially. And, as argued
before, such a $\Sigma $ may be appropriately realized, e.g., by the surface of
constant matter density in the cosmic fluid frame.

\bigskip 
{\sl The lapse function }

\medskip

We postpone to the next section the issue of the $\beta $--evolution of the
rate of volume expansion $K$, since this latter is strictly connected to the
regional averaging of the constraints. On the other hand, for the $\beta $
--evolution of the kinematical variables $(N,N_{b})$ there are physically
sound choices which are directly suggested by the nature of the Raychaudhuri
equation, and from the structure of the equations (\ref{par1}), and (\ref
{par2}). To start with, let us observe that since the leading
lapse--dependent inhomogeneity inducing term in (\ref{raych})  is $\Delta _{g}N$
it is natural to smooth the lapse function $N$ by diffusing its
inhomogeneities by means of the scalar heat equation:
\begin{equation}
\frac{\partial }{\partial \beta }N(\beta )=\Delta _{g(\beta )}N(\beta ) 
\;\;\;;\;\;\;N(\beta =0)=N\;\;,
\label{simpleheat}
\end{equation}
where $\Delta _{g(\beta )}$ is the Laplacian of the metric $g_{ab}(\beta )$
evolved by the Ricci flow. This is basically the harmonic map flow for
the map $N(\beta ):\Sigma \rightarrow \mathbb{R}$. Note also that this is
a non--uniformly parabolic initial value problem, because $\Delta _{g(\beta )}$
depends on the $\beta $--varying metric $g_{ab}(\beta )$. Let us assume that,
for $\beta =0$, the given lapse function is such that $0<\varepsilon \leq
N\leq C_{1}$, $\left| \nabla N(\beta =0)\right| \leq C_{2}$, where $C_{1}$
and $C_{2}$ are suitable constants. In other words, the (instantaneous)
acceleration $\nabla_a \ln N$ of the frame associated with the chosen slice $
\Sigma $ is assumed to be bounded. Then, according to the maximum principle
for (\ref{simpleheat}) (\emph{e.g.} see \cite{aubin}), 
we have $0<\varepsilon \leq N(\beta
)\leq C_{1}$, $\left| \nabla N(\beta )\right| \leq C_{2}$ for $\beta \geq 0$.
If the Ricci flow exists for all $\beta \geq 0$, then by looking at the $
\beta $--evolution of $\left| \nabla N(\beta )\right| ^{2}=g^{ab}(\beta
)\nabla _{a}N(\beta )\nabla _{b}N(\beta )$, a long but elementary 
computation provides:
\begin{eqnarray}
\frac{\partial }{\partial \beta }\left| \nabla N(\beta )\right| ^{2}\nonumber \\
\fl =\Delta_{g(\beta )}\left( \left| \nabla N(\beta )\right| ^{2}\right) -2\left|
\nabla ^{2}N(\beta )\right| ^{2} + \nabla^a N \nabla^b N \left( 
\frac{\partial}{\partial\beta} g_{ab} (\beta) + 2 {\cal R}_{ab}\right)  
\;\;,
\end{eqnarray}
which shows that, if we deform the metric along the Ricci flow, then we have:
\begin{equation}
\fl \frac{\partial }{\partial \beta }\left| \nabla N(\beta )\right| ^{2}=\Delta
_{g(\beta )}\left( \left| \nabla N(\beta )\right| ^{2}\right) -2\left|
\nabla ^{2}N(\beta )\right| ^{2}-\frac{2}{3}\left| \nabla N(\beta )\right|
^{2}\left\langle \mathcal{R}\right\rangle _{\Sigma _{\beta }}\;\;.
\end{equation}
It follows that the maximum of 
$\left| \nabla N(\beta )\right| ^{2}$ is weakly
monotonically decreasing as $\beta \rightarrow \infty $; (the apparent 
condition for this to occur is 
$\left\langle \mathcal{R}\right\rangle_{\Sigma _{\beta}}\geq 0$, but this 
condition is actually not necessary, as can be seen by
rescaling the $\beta-$variable to the standard unnormalized Ricci flow). 
Parabolic theory shows that, 
if the Ricci flow is global (with uniformly bounded curvature),
then also $\beta \rightarrow N(\beta )$ exists for all $\beta $ and $\left|
\nabla N(\beta )\right| \rightarrow 0$ as $\beta \rightarrow \infty $.
Physically, such an averaging procedure implies that the frame instantaneous
acceleration associated with the chosen slice $\Sigma $ is smoothly damped,
and up to a normalization we can assume that as $\beta \rightarrow \infty $, 
$N(\beta )\rightarrow 1$ uniformly. Note that if one identifies $\Sigma $
with a surface of constant matter density or constant $4-$velocity potential, respectively,
in the frame of reference comoving
with the flow lines of the (irrotational) cosmic fluid, then an
inhomogeneous $N$ can be directly related to the instantaneous
acceleration of the fluid particles on $\Sigma $, thus $N(\beta )\rightarrow
1$ as $\beta \rightarrow \infty $ implies that the fluid particles are, on $
\Sigma $, in free fall.

\bigskip

{\sl The shift vector and the matter current density}

\medskip

Since $N_{a}$ generates local
diffeomorphisms in the hypersurface $\Sigma $, a procedure for smoothly
averaging out the shift vector field $N_{a}$ cannot be disentangled from the averaging of the
matter current density $J_{a}$ and from the fact that the averaging of the
distortion tensor $\mathcal{D}_{ab}(\beta )$ generates, according to Lott's
factorization theorem, a geometrically induced shift $v_{a}(\beta )$ (see (\ref
{par2})). Along the same lines of the above remarks one may tentatively
assume that $N_{a}$ can be related to the local $3$--velocities of the
fluid particles on $\Sigma $. On a sufficiently large cosmic domain we may put
ourselves initially on $\Sigma $ (\emph{i.e.}, for $
\beta =0$) into a frame with vanishing shift: $N_{a}=0=J_{a}$. Naively, one
would expect that such a situation will persist as we deform the data $
(\Sigma ,g_{ab},\mathcal{D}_{ab})$. However, as already stressed, there is
a diffeomorphism warp generated by the linearized Ricci flow, which, if not
properly taken into account, will manifest itself as a current $J_a$
of the matter on the smoothed data $(\Sigma ,g_{ab}(\beta ),\mathcal{D}
_{ab}(\beta ))$ when $\beta \rightarrow \infty $. The physical origin of
such diff--induced matter current is rather easy to understand. 
Roughly speaking,
what happens is that the points of the manifold $\Sigma $ are moved around as
curvature bumps are ironed out, (the generator of such a motion is the
gradient in scalar curvature $\nabla _{a}{\cal R}$; these are basically the \emph{
Diff--solitons} familiar in the Ricci flow theory). From a Lagrangian point
of view, this \emph{motion} is transferred to the fluid particles labelled
by the corresponding points of $\Sigma $.

Thus, in order to consistently compensate for such a Ricci flow induced
warping, we must assume that for $\beta \geq 0$ we have a non--vanishing shift
$N_{a}(\beta )$ and a corresponding matter current density $J_{a}(\beta )$.
The natural choice that suggests itself is to introduce a $\beta $
--dependent shift $\beta \rightarrow N_{a}(\beta )$, according to
\begin{equation}
N_{a}(\beta )=\widehat{v}_{a}-v_{a}(\beta )\;\;,  \label{choiceshift}
\end{equation}
where $v_{a}(\beta )$ is the flow solution of (\ref{par2}), and $\widehat{v
}_{a}$ is its $\beta \rightarrow \infty $ limiting value given by (\ref
{finshift}). Note that, according to such a choice,
\begin{equation}
N_{a}(\beta =0)=\widehat{v}_{a}\;,\quad N_{a}(\beta \rightarrow \infty )=0\;\;,
\end{equation}
so that the shift $N_{a}(\beta )$ exactly balances the longitudinal shear ${
\pounds }_{v}g_{ab}(\beta )$ generated by the (linearized) Ricci flow for $
0\leq \beta <\infty $. In other words, there is an optimal choice $\widehat{v
}_{a}$ for the $3$--velocity field of the instantaneous observers on $\Sigma $
which allow them to isotropically smooth the data $(\Sigma ,g_{ab},\mathcal{D
}_{ab})$. If we, e.g., identify $\Sigma $ with a surface of constant average matter
density $\left\langle \varrho \right\rangle _{\mathcal{B}_{\beta }}$, then
the Ricci flow induced shift $\beta \rightarrow N_{a}(\beta )$ gives rise to
a fictious material convection current density $\beta \rightarrow
J_{a}(\beta )$ given by
\begin{equation}
J_{a}(\beta ) := \left\langle \varrho \right\rangle _{\mathcal{B}_{\beta}}
N_{a}(\beta )\;\;,  \label{current}
\end{equation}
where $\left\langle \varrho \right\rangle _{\mathcal{B}_{\beta }}$ is the
averaged matter density over the region of interest, (see (\ref{rhoev})).
Observe that $J_{a}(\beta )\rightarrow 0$ uniformly as $\beta \rightarrow
\infty $.

\bigskip

\subsection{Constraints on regional averaging}

\medskip
\noindent
We are now going to study the asymptotic properties of the variables under
the Lagrangian smoothing flows as $\beta \rightarrow \infty $. For this
purpose we have to go first into the constraint equations in order to
establish a link between the actual and the regionally smoothed--out initial
data.

\newpage


{\sl The constraint equations}

\medskip

If we are willing to assume that Einstein's equations hold on the regions $
B_{\beta }(p_{j};r_{0})$ where we are averaging, then besides the integral
constraint of regional mass preservation, we must also require that our
regional averaging procedure is such as to respect the constraints Eqs.~(\ref
{constraints}), at least on the given scale $r_{0}$. Thus, we restrict the
class of possible deformation flows to act within the solution space of the
constraints. In other words, for each $\beta $ the smoothed geometry should
be a candidate for an admissible initial data set of Einstein's equations in
$B_{\beta }(p_{j};r_{0})$.

\bigskip

{\sl The divergence constraint}

\medskip

Let us start by discussing the divergence constraint, which, in terms of the
trace--free distortion tensor $\mathcal{D}_{ab}$ reads:
\begin{equation}
\nabla ^{b}\left[ \frac{1}{2N}\mathcal{D}_{ab}+\frac{1}{2N}{\pounds}_{\vec{N
}}g_{ab}\right] -\frac{2}{3}\nabla _{a}K=8\pi G\,J_{a}\;\;,
\end{equation}
or more explicitly
\begin{equation}
\fl \frac{1}{2N}\left[ \nabla ^{b}\left( \mathcal{D}_{ab}+{\pounds }_{\vec{N}}
g_{ab}\right) -\left( \mathcal{D}_{ab}+{\pounds }_{\vec{N}}g_{ab}\right)
\nabla ^{b}\ln N\right] -\frac{2}{3}\nabla _{a}K=8\pi G\,J_{a}\;\;.
\end{equation}
If we assume that such a constraint holds in the regions $B_{\beta
}(p_{j};r_{0})$ also for the $\beta $--deformed distortion tensor $\mathcal{D}
_{ab}(\beta )$ we formally get:
\begin{eqnarray}
 \label{divcon2}
\frac{1}{2N(\beta )}\left[ \nabla ^{b}\left( \mathcal{D}_{ab}^{\ast }+{
\pounds }_{\vec{N}+\vec v} g_{ab}\right) -\left( \mathcal{D}
_{ab}^{\ast }+{\pounds }_{\vec{N}+\vec v} g_{ab}\right) \nabla
^{b}\ln N\right] _{\beta } \nonumber \\
-\frac{2}{3}\nabla _{a}K(\beta )\;=\;8\pi G\,J_{a}(\beta )\;\;,
\end{eqnarray}
where the notation $[....]_{\beta }$ is a shorthand for the explicit 
$\beta $--dependence of all the quantities within the brackets.

According to the choice (\ref{choiceshift}) for the $\beta $--dependent shift
vector field $N_{a}(\beta )$, (\ref{divcon2}) reduces to
\begin{eqnarray}
\label{formul}
\frac{1}{2N(\beta )}\left[ \nabla ^{b}\left( \mathcal{D}_{ab}^{\ast }+
{\pounds }_{\widehat{v}}g_{ab}\right) -\left( \mathcal{D}_{ab}^{\ast }+{
\pounds }_{\widehat{v}}g_{ab}\right) \nabla ^{b}\ln N\right] _{\beta } \nonumber \\
-\frac{2}{3}\nabla _{a}K(\beta )=8\pi G\,J_{a}(\beta )\;,
\end{eqnarray}
where the matter current $J_{a}(\beta )$ is given by (\ref{current}).

Assuming that the smoothing flow $\beta \rightarrow (g_{ab}(\beta ),
\mathcal{D}_{ab}^{\ast }(\beta ),v_{a}(\beta ))$ is global and that $
g_{ab}(\beta )\rightarrow \overline{g}_{ab}$ is an isolated constant
curvature metric, then $\mathcal{D}_{ab}^{\ast }(\beta )\rightarrow 0$, $
\nabla ^{b}\ln N\rightarrow 0$ , and $J_{a}(\beta )\rightarrow 0$,
uniformly, and (\ref{formul}) reduces to 
\begin{equation}
\overline{\nabla }^{b}\left( {\pounds }_{\widehat{v}}\overline{g}
_{ab}\right) -\frac{2}{3}\partial _{a}\overline{K}=0  \;\;,\label{kaver}
\end{equation}
where $\overline{A}$ is the $\beta \rightarrow \infty $ limiting value of
the given quantity $A$. Thus, the possible anisotropies in the 
gradient of the rate of
volume expansion $\overline{K}$, as seen by the observers executing the
smoothing process in $B_{\beta }(p_{j};r_{0})$, are only due to the
diff--warp generated by the Ricci flow.

\bigskip

{\sl The Hamiltonian constraint}

\medskip

We are now in position of discussing the Hamiltonian constraint. By adopting
the same regional philosophy applied to the divergence constraint, we
assume that it holds in $B_{\beta }(p_{j};r_{0})$ for the $\beta $%
--dependent data on $(\Sigma_{\beta} ,g_{ab}(\beta ))$. Explicitly, in terms of the
trace--free distortion tensor $\mathcal{D}_{ab}$, we get:
\begin{eqnarray}
\mathcal{R}(\beta )+\frac{2}{3}K^{2}(\beta ) \nonumber \\
-\frac{1}{4N^{2}(\beta )}g^{ab}(\beta )g^{cd}(\beta )\left[ \mathcal{D}
_{ac}^{\ast }+{\pounds }_{\vec{N}+\vec v} g_{ac}\right] _{\beta }
\left[ \mathcal{D}_{bd}^{\ast }+{\pounds }_{\vec{N}+\vec v} g_{bd}
\right] _{\beta } \nonumber \\
=\;16\pi G\rho (\beta )+2\Lambda \;\;.
\end{eqnarray}
According to (\ref{choiceshift}) this reduces to 
\begin{eqnarray}
\fl \mathcal{R}(\beta )+\frac{2}{3}K^{2}(\beta )-\frac{1}{4N^{2}(\beta )}
g^{ab}(\beta )g^{cd}(\beta )\left[ \mathcal{D}_{ac}^{\ast }+{\pounds }_{
\widehat{v}}g_{ac}\right] _{\beta }\left[ \mathcal{D}_{bd}^{\ast }+{\pounds }
_{\widehat{v}}g_{bd}\right] _{\beta } \nonumber \\
=\;16\pi G\rho (\beta )+2\Lambda \;,
\end{eqnarray}
which, upon taking the volume--average over $B_{\beta }(p_{j};r_{0})$, yields:
\begin{eqnarray}
\left\langle \mathcal{R}(\beta )\right\rangle _{\mathcal{B}_{\beta }}+\frac{2
}{3}\left\langle K^{2}(\beta )\right\rangle _{\mathcal{B}_{\beta }}
\label{locham} \nonumber \\
-\frac{1}{4}\left\langle \frac{g^{ab}(\beta )g^{cd}(\beta )\left[ \mathcal{D}
_{ac}^{\ast }+{\pounds }_{\widehat{v}}g_{ac}\right] _{\beta }\left[ \mathcal{
D}_{bd}^{\ast }+{\pounds }_{\widehat{v}}g_{bd}\right] _{\beta }}{N^{2}(\beta
)}\right\rangle _{\mathcal{B}_{\beta }}  \nonumber \\
=\;16\pi G\left\langle \mathcal{\rho }(\beta )\right\rangle _{\mathcal{B}
_{\beta }}+2\Lambda \;\;.
\end{eqnarray}
According to (\ref{rhoev}) we have:
\begin{equation}
\left\langle \mathcal{\rho }(\beta )\right\rangle _{\mathcal{B}_{\beta
}}=\left\langle \mathcal{\rho }\right\rangle _{\mathcal{B}%
_{0}}e^{\int_{0}^{\beta }\left[ \left\langle \mathcal{R}(\lambda
)\right\rangle _{\mathcal{B}_{\lambda}}-\left\langle \mathcal{R}(\lambda
)\right\rangle _{\Sigma _{\lambda}}\right] d\lambda }\;\;.
\end{equation}
Since, in the $\beta \rightarrow \infty $ limit $\mathcal{D}_{bd}^{\ast
}(\beta )\rightarrow 0$, $N(\beta )\rightarrow 1$ uniformly (again by
assuming that the Ricci flow metric $g_{ab}(\beta )\rightarrow \overline{g}
_{ab}$ is an isolated constant curvature metric), we can write (\ref{locham}) as:
\begin{eqnarray}
\left\langle \overline{\mathcal{R}}\right\rangle _{\overline{\mathcal{B}}}+
\frac{2}{3}\left\langle \overline{K}^{2}\right\rangle _{\overline{\mathcal{B}
}} \nonumber \\
\fl =\;16\pi G\left\langle \mathcal{\rho }\right\rangle _{\mathcal{B}
_{0}}e^{\int_{0}^{\infty }\left[ \left\langle \mathcal{R}(\beta
)\right\rangle _{\mathcal{B}_{\beta }}-\left\langle \mathcal{R}(\beta
)\right\rangle _{\Sigma _{\beta }}\right] d\beta }+\frac{1}{4}\left\langle 
\overline{g}^{ab}\overline{g}^{cd}{\pounds }_{\widehat{v}}\overline{g}_{ac}{
\pounds }_{\widehat{v}}\overline{g}_{bd}\right\rangle _{\overline{\mathcal{B}
}}+2\Lambda \;\;. 
\end{eqnarray}
Observe that 
\begin{equation}
\frac{1}{4}\left\langle \overline{g}^{ab}\overline{g}^{cd}{\pounds }_{
\widehat{v}}\overline{g}_{ac}{\pounds }_{\widehat{v}}\overline{g}
_{bd}\right\rangle _{\overline{\mathcal{B}}} = 2\left\langle 
\overline{\sigma }^{2}\right\rangle _{\overline{\mathcal{B}}}\;\;,
\end{equation}
where $\overline{\sigma }^{2}=\frac{1}{2}\overline{\sigma }_{ab}\overline{\sigma }
^{ab} $ is the squared norm of the shear tensor. Moreover, since $
\overline{g}_{ab}$ is a constant curvature metric, $\overline{\mathcal{R}}$
is a constant that, in order to emphasize the regional nature of the
averaging, we denote by $\overline{\mathcal{R}}_{\overline{\mathcal{B}}}$.
Thus, we finally get:
\begin{equation}
\fl \frac{2}{3}\left\langle \overline{K}^{2}\right\rangle _{\overline{\mathcal{B}
}}=16\pi G\left\langle \mathcal{\rho }\right\rangle _{\mathcal{B}
_{0}}e^{\int_{0}^{\infty }\left[ \left\langle \mathcal{R}(\beta
)\right\rangle _{\mathcal{B}_{\beta }}-\left\langle \mathcal{R}(\beta
)\right\rangle _{\Sigma _{\beta }}\right] d\beta}+ 2\left\langle 
\overline{\sigma }^{2}\right\rangle _{\overline{\mathcal{B}}}+2\Lambda -
\overline{\mathcal{R}}_{\overline{\mathcal{B}}}\;\;.  \label{altra}
\end{equation}

\newpage


\subsection{Summary of key--results} 

Before we discuss cosmological implications, we here summarize the key--results of the 
previous sections that are relevant for applications.

In Section~1 we have introduced the concept of a (position--dependent) 
system of geodesic ball--domains on which volume--averages of scalar functions 
are evaluated. We have explicitly shown how the averaged density of matter and the 
geometry of balls change under variation of scale. Complementary to this
(Eulerian) averaging under variation of the ball radius, we have devised a
(Lagrangian) smoothing flow that provides a conceptually equivalent
averaging procedure. Here, we gave substance to the choice of the Ricci
flow as a natural candidate for the smoothing of the metric.
The key--results of this section were:

\begin{itemize}

\item{For scalar functions on a Riemannian 3--manifold the operations
{\em spatial averaging} and {\em rescaling the domain of averaging} do not
commute. In particular, this shows that 
averaging of (scalar) inhomogeneous fields implies the necessity of 
simultaneously rescaling the (tensorial) geometry.}

\item{The metric in the neighborhood of the domain of averaging
is forced to rescale {\em in the direction of its Ricci tensor}.}

\end{itemize}

Important equations were Eq.~(\ref{commutation}) (Proposition 1) and Eq.~(\ref{radscale})
in the context of (Eulerian) averaging under variation of the ball radius, 
and the corresponding equations in the context of (Lagrangian) smoothing,
Eq.~(\ref{geodef}) and Eq.~(\ref{riccifl}).

\medskip

\noindent
In Section~2 we have put into practice the smoothing program in terms of a 
globally volume--preserving Ricci deformation flow. We stressed that the 
choice of this global normalization is technical, the 
key--results concerning regional averages do not depend on this choice.   
We devised a corresponding deformation flow for the material mass under the
assumption that the total material mass within the domain of averaging 
is preserved during the deformation. We then showed that the 
stability properties of the Ricci flow entail a unique choice of the 
smoothing flow for the second fundamental form. We implemented 
the requirement of the preservation of the constraints under such 
deformations, and determined the optimal choice for the reference frame
in which fundamental observers execute the smoothing procedure. 
The key--results of this section were:

\begin{itemize}

\item{The Ricci flow for the metric and the corresponding material mass flow
link the global and regional averages for the material mass density and the
scalar curvature.} 

\item{The second fundamental form must necessarily be deformed according to 
the linearized Ricci flow; there are no other consistent choices.}

\item{The optimal choice of the smoothing reference frame is determined by 
compensating for the fictious motion of fundamental observers that is induced by
the geometrical deformation.}

\item{To get a consistent picture, the constraint equations were required to be preserved 
during the smoothing procedure, i.e., the smoothing flows were required to act within the
solution space of Einstein's equations.}

\end{itemize}

Important equations were Eq.~(\ref{mflow}), Eq.~(\ref{averagedensity}) and Eq.~(\ref{rhoev})
for the Ricci-- and material mass flows, and Eq.~(\ref{trless}) together with Proposition 2
for the linearized Ricci flow in connection with the smoothing of the second fundamental form.
Instead of smoothing the second fundamental form, smoothing of the trace--free distortion 
tensor was suggested, Eq.~(\ref{distortiontensor}). 
Eq.~(\ref{simpleheat}) and Eq.~(\ref{choiceshift}) 
determine the smoothing of the lapse function and shift vector, respectively.
As for the constraints we note that, while the momentum constraint was of technical
importance concerning the consistency of our choice of smoothing the shift vector field,
Eq.~(\ref{formul}), 
(remember that the shift vector field cannot be disentangled from the averaging of the
matter current density, and the averaging of the
distortion tensor $\mathcal{D}_{ab}(\beta )$ generates, according to Lott's
factorization theorem, a geometrically induced shift), 
the Hamiltonian constraint is essential for the following applications, so we
especially point out Eq.~(\ref{locham}) and Eq.~(\ref{altra}).

\medskip

In the next section we are going to discuss average characteristics in 
the smoothed--out region. In particular, we shall 
discuss the effect of averaging and scaling, respectively, on the parameters
of an averaged inhomogeneous cosmology with the key result:

\begin{itemize}

\item{Cosmological parameters, as they are interpreted in a smoothed
cosmology, are `dressed' by the removed geometrical inhomogeneities.}

\end{itemize}

Important equations will be Eq.~(\ref{regionalcurvature}) together with 
Eq.~(\ref{Qregional}) featuring a novel ``curvature backreaction'' effect;
Eq.~(\ref{friedmannregional}) as compared to Eq.~(\ref{genfried2}) for 
generalizations of Friedmann's equation in the actual cosmological model and
the smoothed model, respectively. The Hamiltonian constraint may be cast into 
a constraint equation for {\em regional cosmological parameters} in the 
actual model (Eq.~(\ref{omegaconstraint})) and the smoothed model 
(Eq.~(\ref{omegaconstraintdressed})), which lie at the basis of our interpretation
concerning `bare' and `dressed' cosmological parameters.

\newpage


\section{Cosmological Implications}

We are now going to study Eq.~(\ref{altra}) in a cosmological setting,
addressing especially the role of the cosmological parameters. Before we
discuss them, let us first study and estimate the fluctuations in the rate
of expansion and the scalar curvature.

\subsection{Estimating fluctuations}

In order to write (\ref{altra}) in a form suggesting a generalized Friedmann
equation, let us first introduce the regional variance of the 
distribution of $%
\overline{K}$ in $\overline{\cal B}(p_{j};r_{0})$ measuring the spatial 
fluctuations in the rate of expansion:
\begin{equation}
\delta _{\overline{K}}^{2} := \left\langle \left( \overline{K}-\left\langle 
\overline{K}\right\rangle _{\overline{\mathcal{B}}}\right) ^{2}\right\rangle
_{\overline{\mathcal{B}}}=\left\langle \overline{K}^{2}\right\rangle _{%
\overline{\mathcal{B}}}-\left\langle \overline{K}\right\rangle _{\overline{%
\mathcal{B}}}^{2}\;\;.
\end{equation}
The second step is to exploit (\ref{kaver}) in order to get an estimate
for $\delta _{\overline{K}}^{2}$ in $\overline{\cal B} (p_{j};r_{0})$.
Note that the following estimates are just indicative for the sake of 
understanding; we shall comment on strategies for full estimates later. 

\noindent
Recall that for any ($C^{2}$) function on 
$\overline{\cal B}(p_{j};r_{0})$ we can write (as long as 
${\cal R} (p_{j}) r_0^2 \ll 1$):
\begin{equation}
\fl \frac{1}{V_{E}r_{0}^{3}}\int_{\overline{\cal B}(p_{j};r_{0})} f \;
d\mu _{\overline{g}}=f(p_{j})+\frac{r_{0}^{2}}{10}\left[ 
\Delta f(p_{j})-\frac{\overline{%
\mathcal{R}}_{\overline{\mathcal{B}}}}{3}f(p_{j})\right] + o (r_{0}^{2})
\;\;,
\label{normest}
\end{equation}
where $V_{E}$ is the Euclidean volume of the unit ball in $\mathbb{R}%
^{3}$, and $\Delta $ is the Euclidean Laplacian. Upon applying such a normal
coordinates estimate to $f=\overline{K}^{2}$ and $f=\overline{K}$, 
we obtain for the fluctuations of the rate of expansion:
\begin{equation}
\delta _{\overline{K}}^{2}=\frac{r_{0}^{2}}{10}\left[ 2\left( \delta ^{ab}\partial _{a}%
\overline{K}\partial _{b}\overline{K}\right) (p_{j})+\frac{\overline{%
\mathcal{R}}_{\overline{\mathcal{B}}}}{3}\overline{K}^{2}(p_{j})\right]
+ o (r_{0}^{2})\;\;,
\end{equation}
which, upon substituting the expression (\ref{kaver}) for $\partial _{a}%
\overline{K}$, yields:
\begin{equation}
\delta_{\overline{K}}^{2}=\frac{r_{0}^{2}}{10}\left[ \frac{9}{2}\left( \delta
^{ab}\nabla ^{c}\overline{\sigma _{\shortparallel }}_{ac}\nabla ^{d}%
\overline{\sigma _{\shortparallel }}_{bd}\right) (p_{j})+\frac{\overline{%
\mathcal{R}}_{\overline{\mathcal{B}}}}{3}\overline{K}^{2}(p_{j})\right]
+ o (r_{0}^{2})\;\;,  \label{Hvarian}
\end{equation}
where $\overline{\sigma _{\shortparallel }}_{ac}={\pounds }_{\widehat{v}}%
\overline{g}_{ac}$ is the residual longitudinal shear.

Inserting (\ref{volev}) into the integral of (\ref{rhoev}) we also get:
\begin{equation}
\fl \left\langle \mathcal{\rho }\right\rangle _{\mathcal{B}_{0}}e^{\int_{0}^{%
\infty }\left[ \left\langle \mathcal{R}(\beta )\right\rangle _{\mathcal{B}%
_{\beta }}-\left\langle \mathcal{R}(\beta )\right\rangle _{\Sigma _{\beta }}%
\right] d\beta }=\left\langle \mathcal{\rho }\right\rangle _{\mathcal{B}_{0}}%
\frac{V_{\mathcal{B}_{\beta }}(\beta =0)}{V_{\mathcal{B}_{\beta }}(\beta
\rightarrow \infty )}=\left\langle \mathcal{\rho }\right\rangle _{\mathcal{B}%
_{0}}\frac{V_{\mathcal{B}_{0}}}{V_{\overline{\mathcal{B}}}}=\left\langle 
\mathcal{\rho }\right\rangle _{\overline{\mathcal{B}}}\;.
\end{equation}
A normal coordinates estimate as in (\ref{normest}) (set $f \equiv 1$)
explicitly provides:
\begin{equation}
\left\langle \mathcal{\rho }\right\rangle _{\mathcal{B}_{0}}=\left\langle 
\mathcal{\rho }\right\rangle _{\overline{\mathcal{B}}}\frac{1-\frac{1}{30}
\overline{\mathcal{R}}_{\overline{\mathcal{B}}}r_{0}^{2}}{1-\frac{1}{30}
\mathcal{R}_{\overline{\mathcal{B}}}(p_{j})r_{0}^{2}}+ o (r_{0}^{2})\;,
\end{equation}
revealing the actual dependence of $\left\langle \mathcal{\rho }
\right\rangle _{\mathcal{B}_{0}}$ on the local curvature $%
\mathcal{R}_{\overline{\mathcal{B}}}(p_{j})$ with respect to the regional average 
curvature $\overline{\mathcal{R}}_{\overline{%
\mathcal{B}}}$. In this connection, it is also worthwhile to discuss how
the regional curvature in the smoothed region 
$\overline{\mathcal{R}}_{\overline{%
\mathcal{B}}}$ is related to the \emph{actual} average curvature
$\left\langle \mathcal{R}\right\rangle _{\mathcal{B}_{0}}$. 

Let us start by noting that the (normalized) $\beta-$evolution of the scalar
curvature obeys the following equation \cite{aubin}, \cite{hamilton:ricciflow2}:
\begin{equation}
\frac{\partial {\cal R}(\beta)}{\partial\beta} = \Delta_{g (\beta)}
{\cal R}(\beta)
+ 2 {\cal W}(\beta) - \frac{2}{3}
{\cal R}(\beta)\left\langle {\cal R}(\beta)\right\rangle_{\Sigma_{\beta}}\;\;, 
\end{equation}
where we have defined
\begin{eqnarray}
{\cal W}(\beta) : =\; g^{ab}g^{cd}{\cal R}_{ac}(\beta){\cal R}_{bd}(\beta)
\nonumber \\
=\;g^{ab}g^{cd}[{\tilde{\cal R}}_{ac} + \frac{1}{3}g_{ac}{\cal R}] 
[{\tilde{\cal R}}_{bd} + \frac{1}{3} g_{bd}{\cal R}] 
= {\tilde{\cal R}}^{ab}{\tilde{\cal R}}_{ab} + \frac{1}{3}{\cal R}^2 \;.
\end{eqnarray}
A straightforward calculation then provides:
\begin{eqnarray}
\label{curvatureevolution}
\fl \frac{\partial}{\partial\beta}\langle {\cal R}(\beta)\rangle_{{\cal B}_{\beta}}
= \frac{1}{V_{{\cal B}_{\beta}}}\int_{\partial {\cal B}_{\beta}} 
\nabla_a {\cal R}(\beta) \;d\sigma^a_{g(\beta)} \;\; +\; 
2 \langle {\tilde{\cal R}}^{ab}(\beta)
{\tilde{\cal R}}_{ab}(\beta)\rangle_{{\cal B}_{\beta}}\nonumber\\
\fl - \frac{1}{3} \langle \left({\cal R}(\beta) - \langle {\cal R}(\beta)
\rangle_{{\cal B}_{\beta}}\right)^2 \rangle_{{\cal B}_{\beta}} 
+ \frac{2}{3}
\langle {\cal R}(\beta)\rangle_{{\cal B}_{\beta}} \left(
\langle {\cal R}(\beta)\rangle_{{\cal B}_{\beta}} - \langle {\cal R}(\beta)
\rangle_{\Sigma_{\beta}}\right)\;. 
\end{eqnarray}
On the closed manifold $\Sigma$, {\em i.e.}, taking 
${\cal B}_{\beta} \;\rightarrow\;\Sigma_{\beta}$, 
the first term on the r.h.s. of Eq.~(\ref{curvatureevolution}) reduces
to a flux through an empty boundary and the last term, specifying the 
deviation from the regional to the global average curvature, vanishes,
and Eq.~(\ref{curvatureevolution}) may be integrated to yield:
\begin{equation}
\label{globalcurvature}
\fl \overline{\mathcal{R}}_{\Sigma_{\beta \rightarrow \infty}} =
\left\langle \mathcal{R}%
\right\rangle _{\Sigma _{\beta =0}}+\int_{0}^{\infty }\left[ 2\left\langle 
\widetilde{\cal R}^{ab}\widetilde{\cal R}_{ab}\right\rangle_{\Sigma _{\beta }}
-\frac{1}{3}\left( \left\langle \mathcal{R}^{2}\right\rangle _{\Sigma _{\beta
}}-\left\langle \mathcal{R}\right\rangle _{\Sigma _{\beta }}^{2}\right)
\right] d\beta\;,
\end{equation}
which shows that metrical anisotropies (\emph{i.e.} $\widetilde{\cal R}_{ab}\neq
0 $) tend to generate a ``Friedmannian'' curvature $\overline{\mathcal{R%
}}$ that is larger than the actual averaged 
curvature, whereas large fluctuations tend to an underestimation of
$\overline{\mathcal{R}}$ with respect the real
distribution of curvature. This curvature difference is described by a term
that reminds us of the ``kinematical backreaction'' (having a similar form built
from the extrinsic curvature) often discussed in relation to cosmological 
averaging, e.g., in: \cite{carfora:RG}, \cite{buchert:grgdust}, \cite{buchert:onaverage}.
We therefore introduce as a notional shorthand the 
{\em global curvature backreaction}:
\begin{equation}
\fl {\cal Q}^R_{{\Sigma}} : = \int_0^{\infty} 
\left[\frac{1}{3}\langle \left({\cal R}(\beta) - \langle {\cal R}(\beta)
\rangle_{{\Sigma}_{\beta}}\right)^2 \rangle_{{\Sigma}_{\beta}} - 2 \langle
{\tilde{\cal R}}^{ab}(\beta){\tilde{\cal R}}_{ab}(\beta)
\rangle_{{\Sigma}_{\beta}}\right]\;d\beta 
\;\;.
\end{equation}
Notice that this term vanishes for a FLRW space section, so that globally 
Eq.~(\ref{globalcurvature}) compares two constant curvature models. These
models, in general, also differ by their global volume, which is not manifest in this
equation, because we normalized the Ricci flow to be globally volume--preserving.

On a regional domain of averaging, on which we concentrate in this paper, we can
go one step further and use Eq.~(\ref{volev}) to express the difference between the 
global and the regional average curvature. We can then cast
Eq.~(\ref{curvatureevolution}) into the form:
\begin{eqnarray}
\label{regionalcurvatureevolution}
V_{{\cal B}_{\beta}}^{-2/3}(\beta) \frac{\partial}{\partial\beta}\left[
\langle {\cal R}(\beta)\rangle_{{\cal B}_{\beta}}
 V_{{\cal B}_{\beta}}^{2/3}(\beta) \right]
= \frac{1}{V_{{\cal B}_{\beta}}}\int_{\partial {\cal B}_{\beta}} 
\nabla_a {\cal R}(\beta) \;d\sigma^a_{g(\beta)} \;\; + \nonumber \\ 
2 \langle {\tilde{\cal R}}^{ab}(\beta)
{\tilde{\cal R}}_{ab}(\beta)\rangle_{{\cal B}_{\beta}}
- \frac{1}{3} \langle \left({\cal R}(\beta) - \langle {\cal R}(\beta)
\rangle_{{\cal B}_{\beta}}\right)^2 \rangle_{{\cal B}_{\beta}} \;. 
\end{eqnarray}
In what follows we neglect the flux of curvature through the boundary of the
ball, since we think that this term will not be of any observative 
relevance, at least on sufficiently large portions of the Universe.
A formal integration of Eq.~(\ref{regionalcurvatureevolution}) then
provides the desired relation between the 
(constant) regional curvature in the smoothed model and the actual
regional average curvature:
\begin{equation}
\label{regionalcurvature}
\overline{\cal R}_{\overline{\cal B}} = \langle {\cal R} 
\rangle_{{\cal B}_0} 
\left(\frac{V_{{\cal B}_0}}{V_{\overline{\cal B}}}\right)^{2/3}
- {\cal Q}^R_{{\cal B}_0} \;\;, 
\end{equation}
where we have introduced the {\em regional curvature backreaction}:
\begin{equation}
\label{Qregional}
\fl {\cal Q}^R_{{\cal B}_0}:= \int_0^{\infty} \frac{V_{{\cal B}_{\beta}}(\beta)}
{V_{\overline{\cal B}}} \left[ 
\frac{1}{3}\langle \left({\cal R}(\beta) - \langle {\cal R}(\beta)
\rangle_{{\cal B}_{\beta}}\right)^2 \rangle_{{\cal B}_{\beta}} - 2 \langle
{\tilde{\cal R}}^{ab}(\beta){\tilde{\cal R}}_{ab}(\beta) 
\rangle_{{\cal B}_{\beta}}\right]\;d\beta \;.
\end{equation}
The integral Eq.~(\ref{regionalcurvature}) has the merit that it provides a 
transparent separation of the relevant terms: first, the {\em volume effect}, which is
expected in this form by comparing two constant curvature space sections with the same
matter content for which the regional curvature backreaction vanishes 
(remember that a constant curvature space is proportional to the inverse 
square of the radius of curvature, hence the volume--exponent $2/3$); second,  
the {\em curvature backreaction effect} itself, which consists of a bulk 
contribution and a flux contribution through the boundary (that we neglected). 
Both encode the deviations of the scalar curvature from a constant curvature  
model, e.g., a FLRW space section. 

We are now going to relate our findings to suitable {\em cosmological parameters} by 
moving to a notation that is familiar to cosmology and accessible to the interpretation 
of observations. 

\subsection{The generalized Friedmann equation}

In standard cosmology we are used to discuss cosmological parameters that 
are defined on the basis of a homogeneous--isotropic solution of Einstein's or 
Newton's equations for a self--gravitating distribution of matter.
A refinement of the standard model has been suggested recently
(\cite{buchert:average} in Newtonian cosmology, and \cite{buchert:grgdust},
\cite{buchert:grgfluid} in general relativity), where the (global) 
homogeneous values of the relevant variables were replaced by their 
(regional) spatial volume--averages. 
For example, an averaged dust matter model in relativistic cosmology was found
to obey a set of {\em generalized Friedmann equations} \cite{buchert:grgdust}, from
which we only need the averaged Hamiltonian constraint here (see also \cite{carfora:RG}):
\begin{equation}
6 H^2_{{\cal B}_{0}} - 16\pi G
\langle\varrho\rangle_{{\cal B}_{0}} - 2 \Lambda + \langle{\cal
R}\rangle_{{\cal B}_{0}} = -  {\cal Q}^K_{{\cal B}_{0}}\;, 
\label{friedmannregional} 
\end{equation}
where we have defined, on the averaging domain ${\cal B}_0$, 
the {\em regional Hubble parameter} as $1/3$ of the spatially averaged
rate of expansion $\theta : = - K$:
\begin{equation}
H_{{\cal B}_0}: = \frac{1}{3}\langle \theta \rangle_{{\cal B}_0}\;\;.
\label{hubble}
\end{equation} 
This form of the volume--averaged Hamiltonian constraint 
has the merit to isolate an 
explicit source term (the {\em kinematical backreaction}), which quantifies the
deviations of the average model from the standard FLRW model equation.
It is composed of positive--definite fluctuation terms \cite{buchert:grgdust}:
\begin{equation}
\label{eq:Q-GR} 
{\cal Q}^K_{{\cal B}_0}: = 2 \langle II\rangle_{{\cal B}_0} - \frac{2}{3}
\langle I \rangle_{{\cal B}_0}^2 \;=\; 
\frac{2}{3}\langle ( \theta  - \langle \theta 
\rangle_{{\cal B}_{0}})^2 \rangle_{{\cal B}_{0}} - 2 
\langle \sigma^2 \rangle_{{\cal B}_{0}}\;.
\end{equation}
Here, $I:= K = -\theta$ and 
$II: = \frac{1}{2}(K^2 -
K_{ab}K^{ab}) = \frac{1}{3}\theta^2 - \sigma^2$
denote two of the three principal scalar invariants of
the extrinsic curvature; the latter equality features the corresponding 
kinematical invariants expansion rate $\theta$, and rate of shear $\sigma : = 
\sqrt{\frac{1}{2}\sigma_{ab}\sigma^{ab}}$ (for irrotational flows). 

In contrast to the standard FLRW cosmological parameters there are 
four players. In the former there is by definition no kinematical  
backreaction, ${\cal Q}^K_{{\cal B}_0}=0$. 

\bigskip

{\sl Regional cosmological parameters}

\medskip\noindent
Furthermore, in the general model, we may define 
{\em regional cosmological parameters} as the following (scale--dependent) 
functionals \cite{buchert:grgdust}:
\begin{equation}
\label{standardparameters}
\Omega^M_{{\cal B}_0} : = \frac{8\pi G M_{{\cal B}_0}}{3 V_{{\cal B}_0}
H_{{\cal B}_0}^2 }\;\;\; ;\;\;\;
\Omega^{\Lambda}_{{\cal B}_0}:= \frac{\Lambda}{3 H_{{\cal B}_0}^2 }
\;\;\;;\;\;\;
\Omega^{R}_{{\cal B}_0}:= - \frac{\langle{\cal R}\rangle_{{\cal B}_0}}
{6 H_{{\cal B}_0}^2}\;\;,
\end{equation}
and, in addition to the standard parameters (\ref{standardparameters}),
\begin{equation}
\Omega^{{\cal Q}^K}_{{\cal B}_0} := - \frac{{\cal Q}^K_{{\cal B}_0}}
{6 H_{{\cal B}_0}^2}\;\;.
\end{equation}
These ``parameters'' obey by construction:
\begin{equation}
\Omega^M_{{\cal B}_0} \;+\; \Omega^{\Lambda}_{{\cal B}_0}\;+\; 
\Omega^R_{{\cal B}_0}\;+\; \Omega^{{\cal Q}^K}_{{\cal B}_0} \;=\; 1 
\;\;,
\label{omegaconstraint}
\end{equation}
and they would all become $\beta-$dependent functions under the 
smoothing flow.
Eq.~(\ref{omegaconstraint}) furnishes a way of writing 
the volume--averaged Hamiltonian constraint that is best accessible
to observational interpretations. 

\bigskip

However, unlike in Newtonian cosmology, where the corresponding equations have 
(apart from the definitions of the curvature and backreaction parameters)
a similar form, it is not straightforward to compare the above relativistic
average model parameters to observational parameters.
The reason is that the volume--averages contain information on the 
actually present inhomogeneities in the geometry within the averaging domain.
In contrast, the ``observer's Universe'' is a Euclidean or constant curvature
model\footnote{The stage of interpreting observations is in many cases a 
Newtonian cosmology. In standard cosmology it is common practice to introduce a frame 
that is comoving with a global Hubble flow; in that frame the constant curvature 
is a parameter in the {\em background} FLRW model and the inhomogeneities are 
studied within a Euclidean space section.}.   
Consequently, the interpretation of observations within the set of 
the standard model parameters neglects the geometrical inhomogeneities that
(through the Riemannian volume--average) are hidden in the average
parameters of the realistic cosmology. 

Notwithstanding, we are now in position to relate the parameters interpreted
within the standard model to the actual parameters by studying the 
smoothed cosmological model furnished by Eq.~(\ref{altra}), which can be 
written in terms of the effective quantities 
${\overline{H}}_{\overline{\cal B}}$, $\left\langle \varrho
\right\rangle _{\overline{\cal B}}$, and ${\overline{\cal R}}_{\overline{\cal B}}$
as follows:
\begin{equation}
\label{genfried2}
6 {\overline{H}}^2_{\overline{\cal B}} - 16\pi G
\langle\varrho\rangle_{\overline{\cal B}} - 2 \Lambda + {\overline{\cal
R}}_{\overline{\cal B}} = - {\overline{\cal Q}}^K_{\overline{\cal B}}\;,
\end{equation}
where we have defined the residual kinematical backreaction (after smoothing) by:
\begin{equation}
{\overline{\cal Q}}^K_{\overline{\cal B}}:= 
\frac{2}{3}\delta_{\overline{K}}^{2} 
- 2\left\langle \overline{\sigma }^{2}\right\rangle_{\overline{\cal B}}
\;\;.
\end{equation}
Thus, a ``Friedmannian bias'' in modelling the real (observed) region of the
Universe with a smooth matter distribution evolving in a homogeneous and
isotropic geometry, inevitably `dresses' the matter density $%
\left\langle \varrho\right\rangle_{\overline{\cal B}}$, the
Hubble parameter ${\overline{H}}_{\overline{\cal B}}$, and 
the curvature ${\overline{\cal R}}_{\overline{\cal B}}$
with correction factors, even if the kinematical backreaction effect is respected.
(Note that the latter is expected on a regional domain due to cosmic 
variance of the variables; see our discussion below). 

Correspondingly, an observer with a ``Friedmannian bias'' would interprete
his measurements in terms of the `dressed' cosmological parameters:
\begin{equation}
\fl \overline{\Omega}^M_{\overline{\cal B}} : = 
\frac{8\pi G M_{\overline{\cal B}}}{3 V_{\overline{\cal B}}
\overline{H}_{\overline{\cal B}}^2 }\;\;\; ;\;\;\;
\overline{\Omega}^{\Lambda}_{\overline{\cal B}}:= 
\frac{\Lambda}{3 \overline{H}_{\overline{\cal B}}^2 }
\;\;\;;\;\;\;
\overline{\Omega}^{R}_{\overline{\cal B}}:= 
- \frac{\overline{\cal R}_{\overline{\cal B}}}
{6 \overline{H}_{\overline{\cal B}}^2}\;\;\;;\;\;\;
\overline{\Omega}^{{\cal Q}^K}_{\overline{\cal B}} := 
- \frac{\overline{\cal Q}^K_{\overline{\cal B}}}
{6 \overline{H}_{\overline{\cal B}}^2}\;\;,
\end{equation}
which again, by construction, obey:
\begin{equation}
\overline{\Omega}^M_{\overline{\cal B}} \;+\; 
\overline{\Omega}^{\Lambda}_{\overline{\cal B}}\;+\; 
\overline{\Omega}^R_{\overline{\cal B}}\;+\; 
\overline{\Omega}^{{\cal Q}^K}_{\overline{\cal B}} \;=\; 1 
\;\;.
\label{omegaconstraintdressed}
\end{equation}

\medskip\noindent
Our subsequent analysis will be focussed on discussing the actual
relevance of the geometrical correction terms.

\bigskip

{\sl The relation between `bare' and `dressed' cosmological parameters}

\medskip\noindent
Following from our previous analysis, especially of the expansion and 
curvature fluctuations, we can collect the formulae in order to relate
the `bare' and `dressed' cosmological parameters.

\medskip
Defining the fraction between the volume of the smoothed constant curvature
region and the volume of the original bumby region, as well as the 
fraction of the corresponding Hubble parameters, by
\begin{equation}
\nu : = \frac{V_{\overline{\cal B}}}{V_{{\cal B}_0}}\;\;\;;\;\;\;
\alpha : = \frac{\overline{H}^2_{\overline{\cal B}}}{H^2_{{\cal B}_0}}\;\;,
\end{equation}
we can formally write\footnote{The denominators have to be nonzero; 
degenerate cases must be treated differently. Note that, e.g., the regional 
average curvature is in generic situations nonzero.}:
\begin{equation}
\label{cosmologicalfractions}
\fl \frac{\Omega^M_{{\cal B}_0}}{\overline{\Omega}^M_{\overline{\cal B}}} =
\alpha \; \nu \;\;;\;\;\frac{\Omega^{\Lambda}_{{\cal B}_0}}
{\overline{\Omega}^{\Lambda}_{\overline{\cal B}}} =
\alpha \;;\;
\frac{\Omega^R_{{\cal B}_0}}{\overline{\Omega}^R_{\overline{\cal B}}} =
\alpha \frac{\langle {\cal R}\rangle_{{\cal B}_0}}
{\overline{\cal R}_{\overline{\cal B}}} = \alpha\,\nu^{2/3}\,(1+\mu ) \;;\;
\frac{\Omega^{{\cal Q}^K}_{{\cal B}_0}}
{\overline{\Omega}^{{\cal Q}^K}_{\overline{\cal B}}} =
\alpha \frac{{\cal Q}^K_{{\cal B}_0}}
{{\overline{\cal Q}}^K_{\overline{\cal B}}} \;,
\end{equation}
where in the last equation for the fraction of the curvature parameters we
introduced the dimensionless {\em regional curvature backreaction parameter}
$\mu : = {\cal Q}^R_{{\cal B}_0} / {\overline{\cal R}}_{\overline{\cal B}}$.

\bigskip

This set of equations furnishes a formal basis within which the 
results of this paper can be interpreted with respect to their relevance
for observational cosmology. In the following discussion we comment on
possible strategies for a quantitative analysis of the results.

\subsection{Discussion}

The above listed relations appear to provide a formal recipee to apply the results of
this paper. However, it is clear that a quantitative estimate of the relations
between `bare' and `dressed' cosmological parameters must be based on dynamical 
models and cannot follow solely from geometrical consequences of the smoothing
procedure on a given spatial hypersurface. 

Dynamical estimates can be subtle in a relativistic setting, since realistic models
for the evolution of structure are well--implemented only in the Newtonian framework. 
Recently, progress has been made in estimating the {\em kinematical backreaction}
parameter $\Omega^{{\cal Q}^K}_{{\cal B}_0}$ in Newtonian cosmology. Putting the  key--results
obtained by a realistic Newtonian model for the evolution of structures into 
perspective (\cite{buchert:bks}, \cite{abundance}), it was found that, 
e.g., on a sufficiently large expanding region (of the order of several hundreds 
of Megaparsecs), the kinematical backreaction parameter is quantitatively small,
which is in conformity with other (including relativistic) estimates 
(\cite{futamase1}, \cite{bildhauerfutamase1},
\cite{bildhauerfutamase2}, \cite{seljakhui}, \cite{futamase2},
\cite{russetal:backreaction}). The surprising result, however, was that backreaction
can have a large influence on the other (standard) cosmological parameters during the
dynamical evolution. Although the Newtonian model requires that the backreaction 
vanishes on the global boundary, we may argue in the context of the present work that,
on a given space section at a fixed time of observation, the kinematical backreaction is 
quantitatively less important than the standard parameters, and reflects {\em cosmic
variance} of the measured variables, the presence of which is expected on a regional 
domain of the Universe. Especially the work \cite{buchert:bks} has clearly shown,
however, that the values of these parameters are not related to their initial values
evolved by a FLRW cosmology.

According to our analysis of the effect of smoothing, we found that a similar term
can be identified: the {\em curvature backreaction}, which we expect to play 
an analogous role as the kinematical backreaction.         
In line of these thoughts we therefore suggest to concentrate further quantitative 
investigations on the following, albeit at this level formal, considerations.

Observe that together with the two generalized Friedmann equations
in the form (\ref{omegaconstraint}) and (\ref{omegaconstraintdressed})
for the average (`bare') observables in the real manifold, and for the 
`dressed' observables in the smooth constant curvature model,
\begin{eqnarray}
\Omega^M_{{\cal B}_0} \;+\; \Omega^{\Lambda}_{{\cal B}_0}\;+\; 
\Omega^R_{{\cal B}_0}\;+\; \Omega^{{\cal Q}^K}_{{\cal B}_0} \;=\; 1 \;\;\;\;\;\;
\nonumber\\
\overline{\Omega}^M_{\overline{\cal B}} \;+\; 
\overline{\Omega}^{\Lambda}_{\overline{\cal B}}\;+\; 
\overline{\Omega}^R_{\overline{\cal B}}\;+\; 
\overline{\Omega}^{{\cal Q}^K}_{\overline{\cal B}} \;=\; 1 \;\;,
\end{eqnarray} 
we may consider fractions of various cosmological parameters in order
to eliminate, say the fraction of the Hubble parameters $\alpha$, and 
conclude on the values of the others. Our goal is to relate observationally 
determined values of the `dressed' 
parameters (i.e., as interpreted with a ``Friedmannian bias'') to the 
actual parameters of the average cosmological model.

According to the above discussion of the effect of the 
kinematical backreaction, we are encouraged to consider a subcase in which the
backreaction parameters are quantitatively negligible. (This is, e.g., also true
when the positive--definite fluctuation terms compensate each other: for the
kinematical backreaction we would consider the compensation of shear fluctuations
and expansion fluctuations; for the curvature backreaction we would think of 
a compensation of fluctuations of metrical anisotropies with curvature amplitude 
fluctuations.) It should be emphasized that the following is an illustration and
does not replace the necessity of a fully dynamical investigation.
Approximating the `bare' and `dressed' kinematical backreaction parameters by zero,
we have:    
\begin{equation}
\frac{\Omega^M_{{\cal B}_0}}{\Omega^R_{{\cal B}_0}} \;=\;
\frac{{\overline{\Omega}}^{M}_{\overline{\cal B}}}
{{\overline{\Omega}}^R_{\overline{\cal B}}}\frac{\nu^{1/3}}{1+\mu}
\;\;\;;\;\;\;
\frac{\Omega^M_{{\cal B}_0}}{\Omega^{\Lambda}_{{\cal B}_0}} \;=\;
\frac{{\overline{\Omega}}^{M}_{\overline{\cal B}}}
{{\overline{\Omega}}^{\Lambda}_{\overline{\cal B}}}\;\nu
\;\;.
\end{equation}
Let us furthermore consider a region of the Universe on scales
of the order of $1$ Gpc, where (also in the same spirit possibly biassed) 
observations of the first doppler peak in the CMB fluctuations at the 
``Friedmannian scale'' $\approx 100$Mpc/h favour an approximately
vanishing average curvature $\overline{\cal R}_{\overline{\cal B}}\approx 0$.
If, again for simplicity, we approximate also the curvature backreaction 
parameter by zero, $\mu \approx 0$, in the sense that there are curvature fluctuations
present, but the two positive--definite parts in the backreaction term compensate each other,
we would have an approximately vanishing average curvature also in the actual cosmological model.
Then, the standard argument requires compensation of the actually observed
matter content (including dark baryonic and possibly dark nonbaryonic matter
components), obeying the commonly agreed upper bound 
${\overline{\Omega}}^M_{\overline{\cal B}} \le 0.3$ with a cosmological term
${\overline{\Omega}}^{\Lambda}_{\overline{\cal B}} \approx 0.7$.
For the `bare' parameters we then obtain $\Omega^M_{{\cal B}_0} /
\Omega^{\Lambda}_{{\cal B}_0} \approx  \frac{0.3}{0.7}\nu$, which 
yields the estimate:
\begin{equation}
\Omega^M_{{\cal B}_0} \approx \frac{\frac{0.3}{0.7}\nu}{1+ \frac{0.3}{0.7}\nu}
 \;\;\;;\;\;\;\Omega^{\Lambda}_{{\cal B}_0}\approx 1 - \Omega^M_{{\cal B}_0}
 \;\;.
\end{equation}
This (certainly oversimplified) example shows that, instead of postulating 
the presence of a large cosmological term, the `bare' mass parameter could still acquire 
values close to one, if `undressed', and if the volume fraction $\nu$ is substantially 
greater than 1. The second relation in Eq.~(\ref{cosmologicalfractions}) then shows, that
the actual Hubble--parameter would be larger than the `dressed' Hubble parameter.
The estimation of $\nu$ itself is intimately connected with the estimation
of curvature fluctuations, and a quantitative (scale--dependent) statement about this 
effect is beyond the scope of this paper: it requires dynamical considerations.

\medskip

A quantitative estimate that gives us an idea of the order of magnitude of such an effect
has been worked out by Hellaby \cite{hellaby:volumematching}. Using spherically symmetric
solutions he compares, on some spatial scale, the volume of a FLRW space section with that given by 
a Lema\^\i tre--Tolman--Bondi solution (see also the alternative comparison of volumes 
suggested in \cite{tanimoto}). Contrary to our approach (that compares the two models
at equal material mass), Hellaby uses ``volume matching'' as proposed by Ellis and Stoeger 
\cite{ellisstoeger} (requiring the volumes to be equal).
However, we expect the estimated effect to be of the same order which, for two realistic 
density profiles of a typical cluster of galaxies (an over--density) and a typical `void' 
(an under--density), yields errors in the range $10-30$ \% for these single objects,
if, e.g., the spatial averages of the density profiles are compared with 
the corresponding (fitted) FLRW parameters.

\medskip

\subsection*{Acknowledgements}

{\small TB would like to thank Akio Hosoya, Kei'ichi Maeda and Herbert Wagner
for valuable discussions.
MC would like to thank Kamilla Piotrkowska for many fruitful discussions
that stimulated the interest in this work.
Both of us are greatful for constructive remarks by the referees.

TB acknowledges support by 
the Tomalla Foundation, Switzerland during visits to the University of Geneva in
1999 and 2000, and the University of Pavia during visits in 2000, 2001 and 2002, 
the second with support by a SISSA grant. Since 1.2.2001 TB is supported by
the Sonderforschungsbereich SFB  375 `Astroparticle physics'. 
MC acknowledges support by the SFB 375 during visits to
Ludwig--Maximilians--Universit\"at, Munich in 1999 and 2000,
and partial support by the Ministero dell' Universita' e della Ricerca 
Scientifica under the PRIN project `The Geometry of Integrable Systems'.} 

\bigskip\bigskip

\section*{Remarks}

\noindent
{\bf Remark 1:} In order to put such qualitative
remarks on a firmer ground, we need to quantify in which sense the domains 
${B(p;r)}_{p\in \Sigma }$ do not differ too much from the standard
Euclidean ball $B_{E}(0;r)$. To this end it is useful to introduce the \emph{
Reifenberg norm of} $(\Sigma ,g_{ab})$ \emph{at the scale} $r$, defined according to 
\begin{equation}
\left\| (\Sigma ,g_{ab})\right\| _{r}:= \frac{1}{r}\,\max_{x\in \Sigma
}\;d_{G-H}\left[ B(p;r),B_{E}(0;r)\right] \;,
\end{equation}
where $d_{G-H}\left[ B(p;r),B_{E}(0;r)\right] $ denotes the Gromov--Hausdorff
distance between $B(p;r)$, and $B_{E}(0;r)$ , (recall that $d_{G-H}\left[
B(p;r),B_{E}(0;r)\right] \leq \epsilon $ if $B(p;r)$, and $B_{E}(0;r)$ can
be isometrically embedded in some metric space where they have Hausdorff
distance $\leq \epsilon $). It is immediate to check that $\left\| 
(\Sigma,g_{ab})
\right\| _{r}\rightarrow 0$ as the scale $r\rightarrow 0$, (simply
because for Riemannian manifolds, the geometry of the geodesic balls
approaches the Euclidean geometry as $r\rightarrow 0$ ). Obviously, we are
interested in estimating $\left\| (\Sigma ,g_{ab})\right\| _{r}$ for finite 
$r$.
As a practical matter, this appears to be a difficult task,\ owing to the
rather abstract definition of $\left\| (\Sigma ,g_{ab})\right\| _{r}$. However,
if we assume that the scalar curvature is bounded below,\emph{\ i.e.}, 
\begin{equation}
\inf_{u}\{{\cal R}_{ab}u^{a}u^{b}:\,g_{ab}u^{a}u^{b}=1\}\geq C>-\infty \;,
\end{equation}
then according to a remarkable result of Colding \cite{colding}, the
Reifenberg norm $\left\| (\Sigma ,g_{ab})\right\| _{r}$ is small, \emph{for 
finite} $r$, 
if and only if the volume of $\{B(p;r)\}_{x\in \Sigma }$ is close to
the volume of the Euclidean ball $B_{E}(0;r)$. This result provides a very
useful practical criterion for controlling a scale--dependent averaging
procedure on $(\Sigma ,g_{ab})$, also because once it is established that $%
\left\| (\Sigma ,g_{ab})\right\| _{r_0}$ is small, say by comparing $V\left(
B(p;r_0)\right) $ with $V\left( B_{E}(0;r_0)\right) $ at a given observative
scale $r_0$, then the smallness of $\left\| (\Sigma ,g_{ab})\right\| _{r}$ 
holds at
all scales $r < r_0$. Thus, if the Reifenberg norm $\left\| (\Sigma 
,g_{ab})\right\| _{r}$
is small, \emph{i.e.}, if the geometry of \ $\{B(p;r)\}_{p\in \Sigma }$ does
not vary too wildly with $p$, then the system
of regional averages $\{\left\langle \varrho \right\rangle _{B(p;r)}\}_{p\in
\Sigma }$ appears as a reliable smoothing, at every scale $r$, of the given
matter distribution $\varrho$. 

\bigskip\noindent
{\bf Remark 2:} Even if polar geodesic coordinates suggest themselves as the
most natural labels for the points of $B(p;r)$, they suffer from the basic
drawback that their domain of definition cannot be a priori estimated and
strongly depends on the local geometry of $(\Sigma ,g_{ab})$. In this
connection, a much better control on the geometry of the balls $B(p;r)$, and
hence on  $\left\langle \varrho \right\rangle _{B(p;r)}$, can be achieved by
labelling the points $\exp _{p}^{-1}(q)\in $\ $B_{E}(0,r)$ with harmonic
coordinates, i.e., a coordinate system $\{z^{i}\}$ such that the coordinate
functions $z^{i}$ are harmonic functions with respect to the Laplacian on
$(\Sigma ,g_{ab})$. We can do this by starting from the given (Cartesian)
normal coordinates ${y^{i}}$, and look for a diffeomorphism on a sufficiently
small Euclidean ball $B_{E}(0,r)\subset  \mathbb{R}^{3}$, 
\begin{equation}
\fl \Phi:B_{E}(0,r)\rightarrow B_{E}(0,r) \;\;\;;\;\;\;
y^{i}=\exp _{p}^{-1}(q)\longmapsto \Phi^k (y^{i}):= z^{k}
\end{equation}
such that 
\begin{equation}
\fl \Delta \Phi ^{k}=\frac{1}{\sqrt{\det (g_{ab})}}\partial _{i}\left( \sqrt{
\det (g_{ab})}g^{ij}\partial _{j}\Phi ^{k}\right) =0 
\;\;\;;\;\;\;\Phi ^{k}|_{\partial B_{E}(0,r)}=Id\;.
\end{equation}
The standard theory of elliptic partial differential equations implies that
the harmonic functions so characterized do form a coordinate system in $%
B_{E}(0,r)$. The important observation is that such harmonic coordinates can
be introduced on balls of an a priori size as soon as the manifold $(\Sigma
,g_{ab})$ has bounded sectional curvature and its injectivity radius is bounded
below.


\section*{References}

\begin{thebibliography}{2002}

\bibitem{aubin}
T. Aubin, {\em Some nonlinear problems in Riemannian Geometry}, Springer Verlag
(1998).

\bibitem{bildhauerfutamase1}
S. Bildhauer and T. Futamase, M.\ N.\ R.\ A.\ S.\ {\bf 249}, 126 (1991).

\bibitem{bildhauerfutamase2}
S. Bildhauer and T. Futamase, G.\ R.\ G.\ {\bf 23}, 1251 (1991).

\bibitem{bruni92a}
M. Bruni, P.~.K.~S. Dunsby and G.~F.~R. Ellis, Ap. J. {\bf 395}, 34 (1992).

\bibitem{bruni92b}
M. Bruni, G.~F.~R. Ellis and P.~K.~S. Dunsby, Class.\ Quant.\ Grav. {\bf 9}, 
921 (1992).

\bibitem{buchert:grgdust}
T. Buchert, G.\ R.\ G.\ {\bf 32}, 105 (2000).

\bibitem{buchert:grgfluid}
T. Buchert, G.\ R.\ G.\ {\bf 33}, 1381 (2001).

\bibitem{buchert:onaverage}
T. Buchert, in: {\em 9th JGRG Meeting, 
Hiroshima 1999}, Y. Eriguchi et al. (eds.), pp. 306--321 (2000).

\bibitem{buchert:average}
T. Buchert and J. Ehlers, Astron.\ Astrophys. {\bf 320}, 1 (1997).

\bibitem{buchert:bks}
T. Buchert, M. Kerscher and C. Sicka, Phys.\ Rev.\ D {\bf 62}, 043525 (2000).

\bibitem{buchert:grgscalarfield}
T. Buchert and G. Veneziano, in preparation.

\bibitem{caochow}
H.~D. Cao and B. Chow, Bull. American Math. Soc. {\bf 36}, pp.\ 59--74 (1999).

\bibitem{carfora:deformation0}
M. Carfora and A. Marzuoli, Phys.\ Rev.\ Lett. {\bf 53}, 2445 (1984).

\bibitem{carfora:deformation1}
M. Carfora and A. Marzuoli, Class.\ Quant.\ Grav. {\bf 5}, 659 (1988).

\bibitem{carfora:deformation2}
M. Carfora, J. Isenberg and M. Jackson, J.\ Diff.\ Geom. {\bf 31}, 249 (1990).

\bibitem{carfora:RG}
M. Carfora and K. Piotrkowska, Phys.\ Rev.\ D {\bf 52}, 4393 (1995).

\bibitem{colding}
T.~H. Colding, Ann. Math. {\bf 145}, 477 (1997).

\bibitem{deturck}
D. DeTurck, J.\ Diff.\ Geom. {\bf 18}, 157 (1983). 

\bibitem{dunsby}
P.~K.~S. Dunsby, M. Bruni and G.~F.~R. Ellis, Ap. J. {\bf 395}, 54 (1992).

\bibitem{ehlers:festschrift}
J. Ehlers, in: {\em On Einstein's Path}, Ed.: A. Harvey, 
pp.\ 189--202, Springer, Berlin (1999).

\bibitem{ehlers:newtonian}
J. Ehlers and T. Buchert, G.\ R.\ G.\ {\bf 29},  733  (1997).

\bibitem{ellis:relativistic}
G.~F.~R. Ellis,  in {\em General Relativity and Gravitation} (D. Reidel
Publishing Co., Dordrecht), pp.\ 215--288 (1994).

\bibitem{ellisbruni}
G.~F.~R. Ellis and M. Bruni, Phys.\ Rev.\ D {\bf 40}, 1804 (1989).

\bibitem{ellisstoeger}
G.~F.~R. Ellis and W. Stoeger, Class.\ Quant.\ Grav. {\bf 4}, 1697 (1987).

\bibitem{futamase1}
T. Futamase, M.\ N.\ R.\ A.\ S.\ {\bf 237}, 187 (1989).

\bibitem{futamase2}
T. Futamase, Phys.\ Rev.\ D {\bf 53}, 681 (1996).

\bibitem{hamilton:ricciflow1}
R.~S. Hamilton, J.\ Diff.\ Geom. {\bf 17}, 255 (1982).

\bibitem{hamilton:ricciflow2}
R.~S. Hamilton, in {\em Surveys in Differential Geometry Vol 2}, 
(Boston, MA: International Press), ed. C.~C. Hsiung and S.~T. Yau, pp.7--136 (1995).

\bibitem{hamilton:ricciflow3}
R.~S. Hamilton, Comm.\ Anal.\ Geom. {\bf 7}, 695 (1999).

\bibitem{hellaby:volumematching}
C. Hellaby, G.\ R.\ G.\ {\bf 20}, 1203 (1988).

\bibitem{hosoya:RG}
O. Iguchi, A. Hosoya and T. Koike, Phys.\ Rev.\ D {\bf 57}, 3340 (1998).

\bibitem{isenbergjackson}
J. Isenberg and M. Jackson, J. Diff. Geom. {\bf 35}, 723 (1992).

\bibitem{israel76}
W. Israel, Ann. \ Phys. {\bf 100}, 310 (1976).

\bibitem{abundance}
M. Kerscher, T. Buchert and T. Futamase, Ap. J. {\bf 558}, L79 (2001).

\bibitem{lott}
J. Lott, Comm. Math. Phys. {\bf 107}, 165 (1986).

\bibitem{mukhanov:backreaction}
V.~F. Mukhanov, L.~R.~W. Abramo and R.~H. Brandenberger, Phys.\ Rev.\ Lett. {\bf 78}, 1624 (1997). 

\bibitem{peebles}
P.~J.~E. Peebles, {\em Principles of Physical Cosmology}, Princeton Univ. Press (1993).

\bibitem{petersen}
P. Petersen, {\em Riemannian Geometry}, Springer Verlag GTM 171 (1997).

\bibitem{russetal:backreaction}
H. Russ, M.~H. Soffel, M. Kasai and G. B\"orner, Phys.\ Rev.\  D {\bf 56},
2044 (1997).

\bibitem{seljakhui}
U. Seljak and L. Hui, in {\em Clusters, Lensing and the Future of the
Universe, Proc. 160th Ann. Conf. ASP (University of Maryland, College Park, MD)} Vol.88,
ed. V. Trimble and A. Reisenegger (1995). 

\bibitem{semmes}
S. Semmes, in Appendix of M. Gromov, {\em Metric structures for Riemannian 
and non Riemannian spaces}, Birkh\"auser Boston (1998).

\bibitem{smarryork}
L. Smarr and J.~W. York Jr., Phys.\ Rev.\  D {\bf 17}, 2529 (1978).

\bibitem{sota:RG}
Y. Sota, T. Kobayashi, K. Maeda, T. Kurokawa, M. Morikawa and A. Nakamichi,
Phys.\ Rev.\ D {\bf 58}, 3502 (1998).

\bibitem{tanimoto}
M. Tanimoto, Prog.\ Theor.\ Phys. {\bf 102}, 1001 (1999).

\end{thebibliography}
\end{document}